\documentclass[onecollarge,natbib]{svjour2}
\bibpunct{[}{]}{;}{n}{}{,} 
\smartqed  
\usepackage{graphicx}
\journalname{Few Body Systems}

\newcommand{\lsim}{\mathrel{\rlap{\lower4pt\hbox{\hskip0pt$\sim$}}
\raise1pt\hbox{$<$}}}           
\newcommand{\gsim}{\mathrel{\rlap{\lower4pt\hbox{\hskip0pt$\sim$}}
\raise1pt\hbox{$>$}}}           

\newcounter{dumbone}

\usepackage{color}
\definecolor{purple}{rgb}{0.5,0,0.5}
\definecolor{blue}{rgb}{0.0,0,0.9}

\begin{document}

\title{Masses of ground and excited-state hadrons 
}

\authorrunning{H.\,L.\,L.~Roberts, L.~Chang, I.C.~Clo\"et and C.\,D.~Roberts}
\titlerunning{Masses of ground- and excited-state hadrons}        
\sloppy

\author{Hannes L.\,L.\ Roberts
\and
        Lei Chang
\and
        Ian C.~Clo\"et
\and
        Craig D.~Roberts
}


\institute{Hannes L.\,L. Roberts \and Lei Chang \and Craig D.~Roberts  (\email{cdroberts@anl.gov})
\at
Physics Division, Argonne National Laboratory, Argonne,
Illinois 60439, USA 
\and
Ian C.~Clo\"et \at
Department of Physics, University of Washington, Seattle WA 98195, USA
\and
Hannes L.\,L. Roberts \at
Physics Department, University of California, Berkeley, California 94720, USA
\and
Lei Chang \at
Department of Physics, Center for High Energy Physics and the State Key Laboratory of Nuclear Physics and Technology, Peking University, Beijing 100871, China
\and
Craig D.\ Roberts \at
Institut f\"ur Kernphysik, Forschungszentrum J\"ulich, D-52425 J\"ulich, Germany
\at
Department of Physics, Center for High Energy Physics and the State Key Laboratory of Nuclear Physics and Technology, Peking University, Beijing 100871, China
}

\date{Version: 21 January 2011}

\maketitle

\begin{abstract}
We present the first Dyson-Schwinger equation calculation of the light hadron spectrum that simultaneously correlates the masses of meson and baryon ground- and excited-states within a single framework.  At the core of our analysis is a symmetry-preserving treatment of a vector-vector contact interaction.  In comparison with relevant quantities the root-mean-square-relative-error$/$degree-of freedom is 13\%.  Notable amongst our results is agreement between the computed baryon masses and the bare masses employed in modern dynamical coupled-channels models of pion-nucleon reactions.  Our analysis provides insight into numerous aspects of baryon structure; e.g., relationships between the nucleon and $\Delta$ masses and those of the dressed-quark and diquark correlations they contain.
\end{abstract}


\section{Introduction}
\label{intro}
Spectroscopy has long been a powerful means by which to expose the nature of, and  interactions between, the constituents of a compound system.  In this context hadron spectroscopy has already produced many surprises.  Most notable, perhaps, being the prediction of hadron substructure and its expression in terms of constituent-quark degrees-of-freedom, which were critical steps in the development of quantum chromodynamics (QCD).

With QCD Nature has given us the sole known example of a strongly-interacting quantum field theory that is defined through degrees-of-freedom which cannot directly be detected.  This empirical fact of \emph{confinement} ensures that QCD is the most interesting and challenging piece of the Standard Model.  It means that building a bridge between QCD and the observed properties of hadrons is one of the key problems for modern science.

In hadron physics the constituent-quark model has hitherto been the most widely applied spectroscopic tool \cite{Capstick:2000qj}; and whilst its weaknesses are emphasized by critics and acknowledged by proponents, it is of continuing value because there is nothing better that is yet providing a bigger picture.  On the other hand, steps are being taken with approaches that can rigorously be connected with QCD.  For example, in the continuation of a more-than thirty-year effort, extensive resources are being invested in the application of numerical simulations of lattice-regularised QCD, with the claim that a spectrum which can reasonably be compared with experiment may soon be within reach \cite{Bulava:2009vr,Engel:2010my}.  Herein we bring a different, continuum perspective to computing the hadron spectrum; i.e., QCD's Dyson-Schwinger equations (DSEs) \cite{Roberts:1994dr}.

The DSEs have been applied extensively to the spectrum and interactions of mesons with masses less than 1\,GeV \cite{Roberts:2000aa,Maris:2003vk,Roberts:2007jh}.  On this domain the rainbow-ladder approximation, which is the leading-order in a systematic and symmetry-preserving truncation scheme \cite{Munczek:1994zz,Bender:1996bb}, is an accurate and well-understood tool \cite{Bender:2002as,Bhagwat:2004hn,Matevosyan:2006bk,Eichmann:2008ae,Fischer:2009jm} that can readily be extended to explain properties of the light neutral pseudoscalar mesons \cite{Bhagwat:2007ha}.  Whilst the rainbow-ladder truncation is also reliable for ground-state heavy-heavy mesons \cite{Bhagwat:2006xi}, in order to make progress with heavy-light mesons it is necessary to employ the essentially nonperturbative truncation scheme introduced in Ref.~\cite{Chang:2009zb}.

A Poincar\'e covariant Faddeev equation was formulated in Ref.\,\cite{Cahill:1988dx}, with a first exploration of its solution in Ref.\,\cite{Burden:1988dt}.  The equation is derived following upon the observation that an interaction which describes mesons also generates diquark correlations in the colour-$\bar 3$ channel \cite{Cahill:1987qr}.  Despite the existence of this tractable truncation of the three-body problem in quantum field theory, the spectrum of baryons has almost escaped analysis via the DSEs.
There are some notable exceptions; e.g., studies of the spectrum of ground-state octet and decuplet baryons using simple interaction kernels \cite{Hanhart:1995tc,Buck:1995ch,Oettel:1998bk}, and an extension to baryons containing a single heavy-quark \cite{Ebert:1996ab}.  An attempt has recently been made to unify the treatment of mesons and baryons through the consistent use of a rainbow-ladder truncation \cite{Eichmann:2008ef}.  However, reliable studies are currently available only for the nucleon's ground-state.
Our analysis is a modest step toward unifying a larger subset of meson and baryon spectra via a single interaction: whilst certainly not the last word, it should serve to illustrate the potential of the DSEs in this connection and provide reliable insights.

In Sect.\ref{sec:2} we formulate the bound-state problem for mesons and diquarks.  Following Refs.\,\cite{GutierrezGuerrero:2010md,Roberts:2010rn}, we employ a vector-vector contact-interaction, which is regularised such that confinement is manifest, and chiral symmetry and the pattern by which it is broken are veraciously represented.  Section~\ref{twobodyspectrum} explains how the interaction's three parameters are constrained, and reports a computation of the masses for eight mesons and eight diquark correlations.  Owing to a simplification we employ in constructing the Faddeev equation kernels, two additional parameters appear in our formulation of the nucleon and $\Delta$ Faddeev equations.  This is described in Sect.\,\ref{baryonspectrum}, which also details results for the masses of these states and their parity partners, and the radial excitations of those states.  Section~\ref{epilogue} provides a summary and perspective.

\section{Mesons and diquarks: formulating the bound-state problem}
\label{sec:2}
The bound-state problem for hadrons characterised by two valence-fermions may be studied using the homogeneous Bethe-Salpeter equation
(BSE):
\begin{equation}
[\Gamma(k;P)]_{tu} = \int \! \frac{d^4q}{(2\pi)^4} [\chi(q;P)]_{sr} K_{tu}^{rs}(q,k;P)\,,
\label{genbse}
\end{equation}
where: $\Gamma$ is the bound-state's Bethe-Salpeter amplitude and $\chi(q;P) = S(q+P)\Gamma S(q)$ is its Bethe-Salpeter wave-function; $r,s,t,u$ represent colour, flavour and spinor indices; and $K$ is the relevant fermion-fermion scattering kernel.  This equation possesses solutions on that discrete set of $P^2$-values for which bound-states exist.  (Our Euclidean metric conventions are specified in App.\,\ref{App:EM}.)  In Eq.\,(\ref{genbse}), $S$ is the dressed-quark propagator; viz., the solution of the gap equation:
\begin{equation}
S(p)^{-1}= i\gamma\cdot p + m+ \int \! \frac{d^4q}{(2\pi)^4} g^2 D_{\mu\nu}(p-q) \frac{\lambda^a}{2}\gamma_\mu S(q) \frac{\lambda^a}{2}\Gamma_\nu(q,p) ,
\label{gendse}
\end{equation}
wherein $m$ is the Lagrangian current-quark mass, $D_{\mu\nu}$ is the gluon propagator and $\Gamma_\nu$ is the quark-gluon vertex.

\subsection{Rainbow-ladder truncation}
\label{subsec:RL}
For ground-state, charged pseudoscalar- and vector-mesons constituted from a valence-quark and \mbox{-antiquark} with equal current-mass, the rainbow-ladder truncation of the Bethe-Salpeter and gap equations provides a good approximation \cite{Bender:2002as,Bhagwat:2004hn,Chang:2009zb}.  This means $\Gamma_{\nu}(p,q) =\gamma_{\nu}$ in both Eq.\,(\ref{gendse}) and the construction of $K$ in Eq. (\ref{genbse}), so that one works with
\begin{eqnarray}
S(p)^{-1} & = &i\gamma\cdot p + m+ \int \! \frac{d^4q}{(2\pi)^4} g^2 D_{\mu\nu}(p-q) \frac{\lambda^a}{2}\gamma_\mu S(q) \frac{\lambda^a}{2}\gamma_\nu(q,p) ,
\label{Rdse}\\
\Gamma(k;P) &= & - \int \! \frac{d^4q}{(2\pi)^4} g^2 D_{\mu\nu}(p-q) \frac{\lambda^a}{2}\gamma_\mu S(q+P) \Gamma(q;P)S(q) \frac{\lambda^a}{2}\gamma_\nu \,.
\label{LBSE}
\end{eqnarray}

In this truncation, colour-antitriplet quark-quark correlations (diquarks) are described by an homogeneous Bethe-Salpeter equation that is readily inferred from Eq.\,(\ref{LBSE}); viz. \cite{Cahill:1987qr},
\begin{equation}
\Gamma_{qq}(k;P)H^c = - \int \! \frac{d^4q}{(2\pi)^4} g^2 D_{\mu\nu}(p-q) \frac{\lambda^a}{2}\gamma_\mu S(q+P) \Gamma_{qq}(q;P)H^c[S(-q)]^{\rm T} \left[\frac{\lambda^a}{2}\right]^{\rm T}[\gamma_\nu]^{\rm T} \,,
\label{diquarkBSE}
\end{equation}
where $c=1,2,3$ is a colour label and $\{ H^c \}$ are defined in Eq.\,(\ref{Hmatrices}).  Using the properties of the Dirac and Gell-Mann matrices, it is straightforward to show that
\begin{equation}
\Gamma_{qq}(k;P)C^\dagger = - \frac{1}{2}\int \! \frac{d^4q}{(2\pi)^4} g^2 D_{\mu\nu}(p-q) \frac{\lambda^a}{2}\gamma_\mu S(q+P) \Gamma_{qq}(q;P) C^\dagger S(q) \frac{\lambda^a}{2}\gamma_\nu \,,
\label{LqqBSE}
\end{equation}
which explicates the observation made in the Introduction; i.e., an interaction that binds mesons also generates strong diquark correlations in the colour-$\bar 3$ channel.  It follows moreover that one may obtain the mass and Bethe-Salpeter amplitude for a diquark with spin-parity $J^P$ from the equation for a $J^{-P}$-meson in which the only change is a halving of the interaction strength.  The flipping of the sign in parity occurs because fermions and antifermions have opposite parity.

We note that the rainbow-ladder truncation usually generates asymptotic diquark states.  Such states are not observed and their appearance is an artefact of the truncation.  Higher-order terms in the quark-quark scattering kernel, whose analogue in the quark-antiquark channel do not materially affect the properties of vector and flavour non-singlet pseudoscalar mesons, ensure that QCD's quark-quark scattering matrix does not exhibit singularities which correspond to asymptotic diquark states \cite{Bender:1996bb,Bender:2002as,Bhagwat:2004hn}.   Nevertheless, studies with kernels that don't produce diquark bound states, do support a physical interpretation of the masses, $m_{(qq)_{J^P}}$, obtained using the rainbow-ladder truncation; viz., the quantity $\ell_{(qq)_{J^P}}:=1/m_{(qq)_{J^P}}$ may be interpreted as a range over which the diquark correlation can propagate before fragmentation.

\subsection{Vector-vector contact interaction}
\label{subsec:21}
References \cite{GutierrezGuerrero:2010md,Roberts:2010rn} have shown that a momentum-independent interaction of vector$\,\times\,$vector character is capable of providing a description of $\pi$- and $\rho$-meson static properties which is comparable to that obtained using more-sophisticated, QCD-renormalisation-group-improved interactions \cite{Eichmann:2008ae,Maris:2006ea,Cloet:2007pi}.  This is sufficient justification for proceeding with an analysis of mesons and diquarks using
\begin{equation}
\label{njlgluon}
g^2 D_{\mu \nu}(p-q) = \delta_{\mu \nu} \frac{1}{m_G^2}\,,
\end{equation}
where $m_G$ is a gluon mass-scale. (Such a scale is generated dynamically in QCD \cite{Bowman:2004jm,Cucchieri:2010xr,Aguilar:2010gm,Dudal:2010vn,RodriguezQuintero:2010ss}.)

With this interaction the gap equation becomes
\begin{equation}
 S^{-1}(p) =  i \gamma \cdot p + m +  \frac{4}{3}\frac{1}{m_G^2} \int\!\frac{d^4 q}{(2\pi)^4} \,
\gamma_{\mu} \, S(q) \, \gamma_{\mu}\,.   \label{gap-1}
\end{equation}
The integral possesses a quadratic divergence, even in the chiral limit.  If the divergence is regularised in a Poincar\'e covariant manner, then the solution is
\begin{equation}
\label{genS}
S(p)^{-1} = i \gamma\cdot p + M\,,
\end{equation}
where $M$ is momentum-independent and determined by
\begin{equation}
M = m + \frac{M}{3\pi^2 m_G^2} \int_0^\infty \!ds \, s\, \frac{1}{s+M^2}\,.
\end{equation}

To continue, one must specify a regularisation procedure.  We write \cite{Ebert:1996vx}
\begin{eqnarray}
\nonumber
\frac{1}{s+M^2} & = & \int_0^\infty d\tau\,{\rm e}^{-\tau (s+M^2)}
 \rightarrow \int_{\tau_{\rm uv}^2}^{\tau_{\rm ir}^2} d\tau\,{\rm e}^{-\tau (s+M^2)}
=
\frac{{\rm e}^{- (s+M^2)\tau_{\rm uv}^2}-e^{-(s+M^2) \tau_{\rm ir}^2}}{s+M^2} \,, \label{ExplicitRS}
\end{eqnarray}
where $\tau_{\rm ir,uv}$ are, respectively, infrared and ultraviolet regulators.  It is apparent from Eq.\,(\ref{ExplicitRS}) that a nonzero value of $\tau_{\rm ir}=:1/\Lambda_{\rm ir}$ implements confinement by ensuring the absence of quark production thresholds \cite{Krein:1990sf,Roberts:2007ji}.  Furthermore, since Eq.\,(\ref{njlgluon}) does not define a renormalisable theory,  $\Lambda_{\rm uv}:=1/\tau_{\rm uv}$ cannot be removed but instead plays a dynamical role and sets the scale of all dimensioned quantities.
The gap equation can now be written ($\Gamma(\alpha,y)$ is the incomplete gamma-function)
\begin{equation}
M = m +  \frac{M}{3\pi^2 m_G^2} \,{\cal C}^{\rm iu}(M^2)\,,
\; {\cal C}^{\rm iu}(M^2) = M^2[ \Gamma(-1,M^2 \tau_{\rm uv}^2) - \Gamma(-1,M^2 \tau_{\rm ir}^2)]\,.
\end{equation}

Using the interaction we've specified, the homogeneous BSE for a pseudoscalar meson is
\begin{equation}
\Gamma_{0^-}(P) = - \frac{4}{3} \frac{1}{m_G^2} \,
\int\frac{d^4 q}{(2\pi)^4} \, \gamma_{\mu} S(q+P)\Gamma_{0^-}(P) \gamma_{\mu}  \,. \label{Gamma-eq}
\end{equation}
With a symmetry-preserving regularisation of the interaction in Eq.\,(\ref{njlgluon}), the Bethe-Salpeter amplitude cannot depend on relative momentum and hence may be written
\begin{equation}
\label{genpibsacontact}
\Gamma_{0^-}(P) = \gamma_5 \left[ i E_{0^-}(P) + \frac{1}{M} \gamma\cdot P F_{0^-}(P) \right].
\end{equation}
Crucially, the amplitude contains $F_{0^-}(P)$, a part of pseudovector origin.  It is an essential component of a pseudoscalar meson, which has significant measurable consequences \cite{GutierrezGuerrero:2010md,Roberts:2010rn,Maris:1998hc} and thus cannot be neglected.

Following the discussion in Sect.\,\ref{subsec:RL}, it is straightforward to write the Bethe-Salpeter equation for a $J^P=0^+$ diquark; viz.,
\begin{equation}
\Gamma^C_{qq_{0^+}}(P) = - \frac{2}{3} \frac{1}{m_G^2} \,
\int\frac{d^4 q}{(2\pi)^4} \, \gamma_{\mu} S(q+P) \Gamma^C_{qq_{0^+}}(P) S(q) \gamma_{\mu}  \,,\label{Gamma-qqscalar}
\end{equation}
where
\begin{equation}
\Gamma^C_{qq_{0^+}}(P) = \Gamma_{qq_{0^+}}(P) C^\dagger = \gamma_5 \left[ i E_{qq_{0^+}}(P) + \frac{1}{M} \gamma\cdot P F_{qq_{0^+}}(P) \right].
\end{equation}

\subsection{Ward-Takahashi identities}
In studies of the hadron spectrum it is critical that a computational approach satisfy the vector and axial-vector Ward-Takahashi identities.  Without this it is impossible to preserve the pattern of chiral symmetry breaking in QCD and hence a veracious understanding of hadron mass splittings is not achievable.  The $m=0$ axial-vector identity states ($k_+ = k+P$)
\begin{equation}
\label{avwti}
P_\mu \Gamma_{5\mu}(k_+,k) = S^{-1}(k_+) i \gamma_5 + i \gamma_5 S^{-1}(k)\,,
\end{equation}
where $\Gamma_{5\mu}(k_+,k)$ is the axial-vector vertex, which is determined by
\begin{equation}
\Gamma_{5\mu}(k_+,k) =\gamma_5 \gamma_\mu
- \frac{4}{3}\frac{1}{m_G^2} \int\frac{d^4q}{(2\pi)^4} \, \gamma_\alpha S(q_+) \Gamma_{5\mu}(q_+,q) S(q)\gamma_\alpha\,. \label{aveqn}
\end{equation}
One must therefore implement a regularisation of this inhomogeneous BSE that maintains Eq.\,(\ref{avwti}).

To see what this entails, contract Eq.\,(\ref{aveqn}) with $P_\mu$ and use Eq.\,(\ref{avwti}) within the integrand.  This yields the following two chiral limit identities:
\begin{eqnarray}
\label{Mavwti}
M & = & \frac{8}{3}\frac{M}{m_g^2} \int\! \frac{d^4q}{(2\pi)^4} \left[ \frac{1}{q^2+M^2} +  \frac{1}{q_+^2+M^2}\right],\\
\label{0avwti}
0 & = & \int\! \frac{d^4q}{(2\pi)^4} \left[ \frac{P\cdot q_+}{q_+^2+M^2} -  \frac{P\cdot q}{q^2+M^2}\right]\,,
\end{eqnarray}
which must be satisfied after regularisation.  Analysing the integrands using a Feynman parametrisation, one arrives at the follow identities for $P^2=0=m$:
\begin{eqnarray}
 M & = &  \frac{16}{3}\frac{M}{m_G^2} \int\! \frac{d^4q}{(2\pi)^4} \frac{1}{[q^2+M^2]}, \label{avwtiMc}\\
0 & = & \int\! \frac{d^4q}{(2\pi)^4} \frac{\frac{1}{2} q^2 + M^2 }{[q^2+M^2]^2} \label{avwtiAc}.
\end{eqnarray}

Equation\,(\ref{avwtiMc}) is just the chiral-limit gap equation.  Hence it requires nothing new of the regularisation scheme.   On the other hand, Eq.\,(\ref{avwtiAc}) states that the axial-vector Ward-Takahashi identity is satisfied if, and only if, the model is regularised so as to ensure there are no quadratic or logarithmic divergences.  Unsurprisingly, these are the just the circumstances under which a shift in integration variables is permitted, an operation required in order to prove Eq.\,(\ref{avwti}).

We observe in addition that Eq.\,(\ref{avwti}) is valid for arbitrary $P$.  In fact its corollary, Eq.\,(\ref{Mavwti}), can be used to demonstrate that in the chiral limit the two-flavour scalar-meson rainbow-ladder truncation of the contact-interaction DSEs produces a bound-state with mass $m_\sigma = 2 \,M$ \cite{Roberts:2010gh} (see App.\,\ref{app:scalar}).  The second corollary, Eq.\,(\ref{0avwti}), entails
\begin{equation}
0 = \int_0^1d\alpha \,
\left[ {\cal C}^{\rm iu}(\omega(M^2,\alpha,P^2))  + \, {\cal C}^{\rm iu}_1(\omega(M^2,\alpha,P^2))\right], \label{avwtiP}
\end{equation}
with $\omega(M^2,\alpha,P^2) = M^2 + \alpha(1-\alpha) P^2$ and ${\cal C}^{\rm iu}_1(z) = - z (d/dz){\cal C}^{\rm iu}(z)$.

\section{Mesons and diquarks: computed masses}
\label{twobodyspectrum}
In App.\,\ref{app:groundBSEs} we present explicit forms of the homogeneous rainbow-ladder Bethe-Salpeter equations for ground-state $J^P=0^-$, $0^+$, $1^-$, $1^+$ mesons and diquarks, obtained using the interaction and regularisation scheme described above.  In order to present numerical results, the values of our parameters must be fixed.

\subsection{Ground states}
\begin{table}[t]
\caption{Results obtained with (in GeV) $m_G=0.132\,$, $\Lambda_{\rm ir} = 0.24\,$, $\Lambda_{\rm uv}=0.905$, which yield a root-mean-square relative-error of 13\% in comparison with our specified goals for the observables. Dimensioned quantities are listed in GeV.
\label{Table:static}
}
\begin{center}
\begin{tabular*}
{\hsize}
{
l@{\extracolsep{0ptplus1fil}}
|c@{\extracolsep{0ptplus1fil}}
c@{\extracolsep{0ptplus1fil}}
c@{\extracolsep{0ptplus1fil}}
|c@{\extracolsep{0ptplus1fil}}
c@{\extracolsep{0ptplus1fil}}
c@{\extracolsep{0ptplus1fil}}
c@{\extracolsep{0ptplus1fil}}
c@{\extracolsep{0ptplus1fil}}
c@{\extracolsep{0ptplus1fil}}}\hline
$m$ & $\bar E_\pi$ & $\bar F_\pi$ & $\bar E_\rho$ & $M$ & $\kappa_\pi^{1/3}$ & $m_\pi$ & $m_\rho$ & $f_\pi$ & $f_\rho$\\\hline
0 & 3.568 & 0.459 & 1.520 & 0.358 & 0.241 & 0\,~~~~~ & 0.919 & 0.100 & 0.130\rule{0ex}{2.5ex}\\
0.007 & 3.639 & 0.481 & 1.531 & 0.368 & 0.243 & 0.140 & 0.928 & 0.101 & 0.129
\\\hline
\end{tabular*}
\end{center}
\end{table}

The interaction defined in Sect.\,\ref{subsec:21} possesses three parameters: $\Lambda_{\rm ir}$, $\Lambda_{\rm uv}$, and $m_G$.  We fix $\Lambda_{\rm ir}=0.24\,$GeV\,$ \approx \Lambda_{\rm QCD}$,  since $r_{\rm QCD}:=1/\Lambda_{\rm QCD}\sim 0.8\,$fm is a length-scale typical of confinement, so that only the other two parameters are active.  We fix their values by performing a least-squares fit in the chiral limit to $M^0=0.40\,$GeV, $m_\rho^0= 0.78\,$GeV, $f_\pi^0=0.088\,$GeV, $f_\rho^0 = 0.15\,$GeV $\kappa_\pi^0= (0.22\,{\rm GeV})^3$.  The hitherto undefined entries in this list are: the light-meson leptonic decay constants\footnote{These expressions may be computed in a straightforward manner from the general formulae in, e.g., Ref.\,\protect\cite{Ivanov:1998ms}.}
\begin{equation}
\label{fpirho}
f_\pi = \frac{1}{M}\frac{3}{2\pi^2} \,[ \bar E_\pi - 2 \bar F_\pi] \,{\cal K}_{FE}^\pi(P^2=-m_\pi^2) ,\;
f_\rho = - \frac{9}{2} \, \frac{\bar E_\rho}{m_\rho} \, K^\rho(-m_\rho^2)
\end{equation}
and the in-pion condensate \cite{Brodsky:2010xf}
\begin{equation}
\label{kpim}
\kappa_\pi =  f_\pi \frac{3}{4\pi^2} [ \bar E_\pi {\cal K}_{EE}^\pi(-m_\pi^2) + \bar F_\pi \,{\cal K}_{EF}^\pi(-m_\pi^2) ]\,.
\end{equation}
In these expressions, $\bar E_\pi$, $\bar F_\pi$ and $\bar E_\rho$ are the canonically-normalised Bethe-Salpeter amplitudes (see App.\,\ref{app:groundBSEs}).  This procedure yields the results in Table~\ref{Table:static}, and the masses of the meson and diquark ground states reported in Tables~\ref{Mesonmasses} and \ref{Diquarkmasses}.

\begin{table}[t]
\caption{\label{Mesonmasses}
Meson masses (GeV) computed using our contact-interaction DSE kernel, which produces a momentum-independent dressed-quark mass $M=0.37\,$GeV from a current-quark mass of $m=7\,$MeV.  ``RL'' denotes rainbow-ladder truncation.  Row-2 is obtained by augmenting the RL kernel with repulsion generated by vertex dressing (see Eq.\,(\protect\ref{gSO}) and associated text).  The text around Eq.\,(\protect\ref{zerovalue}) explains the errors on the masses of radially-excited states.  Row-3 lists experimental masses \protect\cite{Nakamura:2010zzi} for comparison.  NB.\ We implement isospin symmetry so, e.g., $m_\omega=m_\rho$, $m_{f_1}=m_{a_1}$, etc.}
\begin{center}
\begin{tabular*}
{\hsize}
{
l@{\extracolsep{0ptplus1fil}}
|c@{\extracolsep{0ptplus1fil}}
c@{\extracolsep{0ptplus1fil}}
c@{\extracolsep{0ptplus1fil}}
c@{\extracolsep{0ptplus1fil}}
|c@{\extracolsep{0ptplus1fil}}
c@{\extracolsep{0ptplus1fil}}
c@{\extracolsep{0ptplus1fil}}
c@{\extracolsep{0ptplus1fil}}}\hline
   & $m_\pi$ & $m_\rho$ & $m_\sigma$ & $m_{a_1}$
   & $m_{\pi^\ast}$ & $m_{\rho^\ast}$ & $m_{\sigma^\ast}$ & $m_{a_1^\ast}$ \\\hline
%
%
RL & 0.14 & 0.93 & 0.74 & 1.08 & $1.38\pm 0.06$ & $1.29\pm 0.07$ & $1.41\pm0.06$ & $1.30\pm 0.06$\rule{0em}{2.5ex}\\
%
%
RL $\ast \;g_{\rm SO}^2$ & 0.14 & 0.93 & 1.29 & 1.38
                    & $1.38\pm 0.06$ & $1.29\pm 0.07$ & $1.47\pm0.04$ & $1.47\pm 0.03$\rule{0em}{2.5ex}\\\hline
experiment & 0.14 & 0.78 & 0.4 -- 1.2 &  1.24 &1.3$\,\pm\,$0.1 & 1.47 & 1.2 -- 1.5 & 1.43\rule{0em}{2.5ex}\\\hline
\end{tabular*}
\end{center}
\end{table}

The pattern of meson masses is typical of the rainbow-ladder truncation \cite{Maris:2006ea,Cloet:2007pi,Watson:2004kd}: $\pi$- and $\rho$-mesons are described well but their parity partners -- the $\sigma$- and $a_1$-mesons -- are not.  The origin and solution of this longstanding puzzle are now available following a novel reformulation of the BSE \cite{Chang:2009zb}, which is valid and tractable when the quark-gluon vertex is fully dressed.  In employing this approach to study the meson spectrum it was found that DCSB generates a large dressed-quark anomalous chromomagnetic moment and consequently that spin-orbit splitting between ground-state mesons is dramatically enhanced \cite{Chang:2010jq,Chang:2010hb}.  This is the mechanism responsible for a magnified splitting between parity partners; namely, essentially-nonperturbative DCSB corrections to the rainbow-ladder truncation largely-cancel in the pseudoscalar and vector channels but add constructively in the scalar and axial-vector channels.

With this in mind, we introduced spin-orbit repulsion into the scalar- and pseudovector-meson channels through the artifice of a phenomenological coupling $g^2_{\rm SO}\leq 1$, introduced as a factor multiplying the kernels defined in Eqs.\,(\ref{a1Kernel}), (\ref{sigmaKernel}).  The value\footnote{NB.\, $g_{\rm SO}=1$ means no repulsion.  The mass changes slowly with diminishing $g_{\rm SO}$; e.g., $g_{\rm SO}=0.50$ yields $m_{a_1}=1.23\,$GeV.}
\begin{equation}
\label{gSO}
g_{\rm SO} = 0.240
\end{equation}
is chosen so as to obtain the experimental value for the $a_1$-$\rho$ mass-splitting, which we know to be achieved by the corrections described above \cite{Chang:2009zb,Chang:2010jq,Chang:2010hb}.  This expedient produces the results in Row-2 of Tables~\ref{Mesonmasses} and \ref{Diquarkmasses}.  It is noteworthy that the shift in $m_{a_1}$ is accompanied by an increase of $m_\sigma$ and that the new value matches an estimate for the $\bar q q$-component of the $\sigma$-meson obtained using unitarised chiral perturbation theory \cite{Pelaez:2006nj}.

Tensor mesons are made conspicuous by their absence from Table~\ref{Mesonmasses}.  This is readily explained.  In constituent-quark models one may only construct a $J=2$ state constituted from two $J=\frac{1}{2}$ quarks if the system's ground state contains at least one unit of orbital angular momentum.  The analogue of this statement in quantum field theory is expressed in the requirement that a normal tensor meson's Bethe-Salpeter amplitude must depend at least linearly on the relative momentum  \cite{LlewellynSmith:1969az}.  This is impossible using a symmetry-preserving regularisation of the interaction in Eq.\,(\ref{njlgluon}) and hence it doesn't generate tensor meson bound states.

\begin{table}[t]
\caption{\label{Diquarkmasses}
Diquark masses (GeV) computed using our contact-interaction DSE kernel, which produces a momentum-independent dressed-quark mass $M=0.37\,$GeV from a current-quark mass of $m=7\,$MeV.  ``RL'' denotes rainbow-ladder truncation.  Row-2 is obtained by augmenting the RL kernel with repulsion generated by vertex dressing (see Eq.\,(\protect\ref{gSO}) and associated text).}
\begin{center}
\begin{tabular*}
{\hsize}
{
l@{\extracolsep{0ptplus1fil}}
|c@{\extracolsep{0ptplus1fil}}
c@{\extracolsep{0ptplus1fil}}
c@{\extracolsep{0ptplus1fil}}
c@{\extracolsep{0ptplus1fil}}
|c@{\extracolsep{0ptplus1fil}}
c@{\extracolsep{0ptplus1fil}}
c@{\extracolsep{0ptplus1fil}}
c@{\extracolsep{0ptplus1fil}}}\hline
   & $m_{qq_{0^+}}$ & $m_{qq_{1^+}}$ & $m_{qq_{0-}}$ & $m_{qq_{1^-}}$
   & $m_{qq_{0^+}^\ast}$ & $m_{qq_{1^+}^\ast}$ & $m_{qq_{0-}^\ast}$ & $m_{qq_{1^-}^\ast}$ \\\hline
%
%
RL & 0.78 & 1.06 & 0.93 & 1.16 & $1.39\pm 0.06$ & $1.32\pm 0.05$ & $1.42\pm0.05$ & $1.33\pm 0.05$\rule{0em}{2.5ex} \\
%
%
RL $ \ast \; g_{\rm SO}^2$ & 0.78 & 1.06 & 1.37 & 1.45
                    & $1.39\pm 0.06$ & $1.32\pm0.05$ & $1.50\pm 0.03$ & $1.52\pm 0.02$\rule{0em}{2.5ex}\\\hline
\end{tabular*}
\end{center}
\end{table}

Some remarks on the spectrum of ground-state diquarks are also appropriate here.  Our computed values for the masses of the scalar and axial-vector diquarks are commensurate with other estimates based on the rainbow-ladder truncation \cite{Burden:1996nh,Maris:2002yu}; and with numerical simulations of lattice-QCD \cite{Alexandrou:2006cq}.  In addition, it is noteworthy that our results for the $g_{\rm SO}$-corrected masses of the ground-state pseudoscalar and vector diquarks are just 10\% smaller than the values determined in Ref.\,\cite{Burden:1996nh}.  Until recently the separable model employed therein for the Bethe-Salpeter kernel was unique in providing a realistic value for the $a_1$-$\rho$ mass-splitting \cite{Bloch:1999vka}.

It is also of interest to elucidate the role of the scalar diquark's vector component; i.e., the $F_{qq_{0^+}}$-term in its Bethe-Salpeter amplitude.  Absent this term, the diquark mass drops by $62\,$MeV.  Its presence therefore produces a small amount of repulsion.  In Ref.\,\cite{Burden:1996nh} this vector-component of the Bethe-Salpeter amplitude was found to provide a repulsive shift of $83\,$MeV in the scalar diquark's mass, whilst in Ref.\,\cite{Maris:2002yu} the shift is $+80\,$MeV.

\subsection{Meson and diquark radial excitations}
\label{mesonradial}
In quantum mechanics the radial wave function for a bound-state's first radial excitation possesses a single zero.  A similar feature is expressed in quantum field theory: namely, in a fully covariant approach a single zero is seen in the relative-momentum dependence of the leading Tchebychev moment of the dominant Dirac structure in the bound state amplitude for a meson's first radial excitation \cite{Holl:2004fr}.  The existence of radial excitations is therefore very obvious evidence against the possibility that the interaction between quarks is momentum-independent: a bound-state amplitude that is independent of the relative momentum cannot exhibit a single zero.  One may also express this differently; namely, if the location of the zero is at $k_0^2$, then a momentum-independent interaction can only produce reliable results for phenomena that probe momentum scales $k^2\ll k_0^2$.  In QCD, $k_0\sim M$ and hence this criterion is equivalent to that noted in Ref.\,\cite{GutierrezGuerrero:2010md}.

Herein, however, we skirt this difficulty by means of an expedient employed in Ref.\,\cite{Volkov:1999xf}; i.e., we insert a zero by hand into the kernels defined in Eqs.\,(\ref{piKEE}) -- (\ref{piKFF}), (\ref{Kgamma}), (\ref{a1Kernel}), (\ref{sigmaKernel}).  This means that we identify the BSE for a radial excitation as the form of Eq.\,(\ref{LBSE}) obtained with Eq.\,(\ref{njlgluon}) and insertion into the integrand of a factor
\begin{equation}
1 - d_{\cal F} q^2\,,
\end{equation}
which forces a zero into the kernel at $q^2=1/d_{\cal F}$, where $d_{\cal F}$ is a parameter.  It is plain that the presence of this zero has the effect of reducing the coupling in the BSE and hence it increases the bound-state's mass.  Although this may not be as transparent with a more sophisticated interaction, a qualitatively equivalent mechanism is always responsible for the elevated values of the masses of radial excitations.

To illustrate our procedure, consider the BSE for the vector meson, in which the following replacement is made:\footnote{The procedure is a little more involved in the pseudoscalar channel owing to the axial-vector Ward-Takahashi identity and the richer structure of the Bethe-Salpeter amplitude.  It is detailed in App.\,\protect\ref{app:groundBSEspion}.}
\begin{equation}
K^\rho(P^2) \longrightarrow K^{\rho^\ast}(P^2)
=\frac{1}{3\pi^2m_G^2} \int_0^1d\alpha\, \alpha(1-\alpha) P^2\,  \overline{\cal F}^{\rm iu}_1(\omega(M^2,\alpha,P^2))\,, \label{Krhoprime}
\end{equation}
where
\begin{eqnarray}
{\cal F}^{\rm iu}(\omega(M^2,\alpha,P^2))
&=& {\cal C}^{\rm iu}(\omega(M^2,\alpha,P^2))
- d_{\cal F} {\cal D}^{\rm iu}(\omega(M^2,\alpha,P^2))\,,\\
{\cal D}^{\rm iu}(\omega(M^2,\alpha,P^2)) & = & \int_0^\infty ds\,s^2\,\frac{1}{s+M^2}
\to  \int_{r_{\rm uv}^2}^{r_{\rm ir}^2} d\tau\, \frac{2}{\tau^3} \,
\exp\left[-\tau \omega(M^2,\alpha,P^2)\right],
\end{eqnarray}
${\cal F}^{\rm iu}_1(z) = - z (d/dz){\cal F}^{\rm iu}(z)$ and $\overline{\cal F}_1(z) = {\cal F}_1(z)/z$.

Regarding the location of the zero, motivated by extant studies in the pseudoscalar channel \cite{Holl:2004fr}, we choose $1/d_{\cal F} = M^2$.  The position of the zero in the leading Tchebychev moment of an excited state in a given channel is an indication of that state's size.  Hence it is an oversimplification to place the zero at the same location in each channel.  Therefore, in Tables~\ref{Mesonmasses} and \ref{Diquarkmasses}, we report results obtained subject to a 20\% variation in the zero's location; i.e., determined with
\begin{equation}
\label{zerovalue}
\frac{1}{d_{\cal F}} = M^2 \, (1.0 \pm 0.2)\,.
\end{equation}

In order to evaluate the credibility of masses we compute subsequently for baryons, it is important to consider critically the results in the right columns of Table~\ref{Mesonmasses}.  In this connection it is noteworthy that whilst no parameters were tuned the computed masses are in good agreement with the known spectrum.  Hence, this simple model produces a phenomenology which represents a considerable improvement over that of existing DSE studies.  We therefore judge it to provide a solid foundation for a study of baryons.

\section{Spectrum of Baryons}
\label{baryonspectrum}
We compute the masses of light-quark baryons using a Faddeev equation built from the interaction in Eq.\,(\ref{njlgluon}) and the diquark correlations discussed quantitatively in Sect.\,\ref{twobodyspectrum}.  The general structure of this equation for nucleon and $\Delta$ states is described in App.\,\ref{app:FEgeneral}.

The bound-state equations specific to our model are defined once the detailed forms of the kernels in Eqs.\,(\ref{FEone}) and (\ref{FEDelta}) are specified; namely, by the structure of the dressed-quark propagator, the diquark Bethe-Salpeter amplitudes and the diquark propagators.
In completing these kernels we make a drastic simplification; viz., in the Faddeev equation for a baryon of type $B=N,\Delta$, the quark exchanged between the diquarks is represented as
\begin{equation}
S^{\rm T}(k) \to \frac{g_B^2}{M}\,,
\label{staticexchange}
\end{equation}
where $g_B$ is discussed below.  This is a variant of the so-called ``static approximation,'' which itself was introduced in Ref.\,\cite{Buck:1992wz} and has subsequently been used in studies of a range of nucleon properties \cite{Bentz:2007zs}.  In combination with diquark correlations generated by Eq.\,(\ref{njlgluon}), whose Bethe-Salpeter amplitudes are momentum-independent, Eq.\,(\ref{staticexchange}) generates Faddeev equation kernels which themselves are momentum-independent.  The dramatic simplifications which this produces are the merit of Eq.\,(\ref{staticexchange}).

\subsection{Ground-state $\Delta$ and nucleon}
\label{DeltaFEexplicit}
Owing to its inherent simplicity, we use the $\Delta$ to illustrate the construction of a Faddeev equation.  With a momentum-independent kernel, the Faddeev amplitude cannot depend on relative momentum.  Hence Eq.\,(\ref{DeltaFA}) becomes
\begin{equation}
{\cal D}_{\nu\rho}(\ell;P) u_\rho(P) = f^\Delta(P) \, \mbox{\boldmath $I$}_{\rm D} \, u_\nu(P)\,.
\label{Dnurho}
\end{equation}
NB.\ Regarding Eq.\,(\ref{DeltaFA}) in general, one might naively suppose that isospin-one tensor diquarks could play a material role in the Faddeev amplitude for a ground state $\Delta$.  However, this notion can quickly be discarded because ground-states are distinguished by containing the smallest amount of quark orbital angular momentum, $L$, and a tensor diquark is characterised by $L\geq 1$.

Using Eq.\,(\ref{Dnurho}), Eq.\,(\ref{FEDelta}) can be written
\begin{equation}
f^\Delta(P) \, u_\mu(P)  = 4 \frac{g_\Delta^2}{M} \int\frac{d^4\ell}{(2\pi)^4}\,{\cal M}^\Delta_{\mu\nu}(\ell;P)\,f^\Delta(P) \, u_\nu(P)\,,
\end{equation}
with ($K=-\ell+P$, $P^2=-m_\Delta^2$)
\begin{equation}
{\cal M}^\Delta_{\mu\nu}(\ell;P) = 2\,i\Gamma_\rho^{1^+}(K) i\bar \Gamma_\mu^{1^+}(-P) S(\ell) \Delta^{1^+}_{\rho\nu}(K)\,,
\end{equation}
where the ``2'' has arisen through the isospin contractions.

At this point, one post-multiplies by $\bar u_\beta(P;r)$ and sums over the polarisation index to obtain, Eq.\,(\ref{Deltacomplete}),
\begin{equation}
\Lambda_+(P) R_{\mu\beta}(P)  = 4 \frac{g_\Delta^2}{M} \int\frac{d^4\ell}{(2\pi)^4}\,{\cal M}^\Delta_{\mu\nu}(\ell;P)\,
\Lambda_+(P) R_{\nu\beta}(P) \,,
\end{equation}
which, after contracting with $\delta_{\mu\beta}$, yields
\begin{eqnarray}
1  &=& \frac{g_\Delta^2}{M} \, {\rm tr}_{\rm D}\int\frac{d^4\ell}{(2\pi)^4}\,{\cal M}^\Delta_{\mu\nu}(\ell;P)\,
\Lambda_+(P) R_{\nu\mu}(P) \\
\nonumber & = & \frac{8}{3} \frac{g_\Delta^2}{M m_\Delta^3} \frac{E_{qq_{1^+}}^2}{m_{qq_{1^+}}^2} \!\int\frac{d^4\ell}{(2\pi)^4}
\frac{1}{(K^2+m_{qq_{1^+}}^2)(\ell^2+M^2)}
\left( -\ell\cdot P \, [3 \, m_{qq_{1^+}}^2 m_\Delta^2 + (K\cdot P)^2] \right.\\
&& \rule{10em}{0ex} \left. +m_\Delta[ 2 m_\Delta \ell\cdot K K\cdot P + 3 M (m_{qq_{1^+}}^2 m_\Delta^2 + (K\cdot P)^2)]\right),
\end{eqnarray}
where $E_{qq_{1^+}}\!(K)$ is the canonically-normalised axial-vector diquark Bethe-Salpeter amplitude, Eq.\,(\ref{CNavdiquark}).  Now, with the aid of a Feynman parametrisation, the right hand side becomes
\begin{eqnarray}
\nonumber
&& \frac{8}{3} \frac{g_\Delta^2}{M m_\Delta^3} \frac{E_{qq_{1^+}}^2}{m_{qq_{1^+}}^2}
\! \int\frac{d^4\ell}{(2\pi)^4} \int_0^1 d\alpha\,
\frac{1}{[(\ell-\alpha P)^2 + \sigma_\Delta(\alpha,M,m_{qq_{1^+}},m_\Delta)]^2}
\left( -\ell\cdot P \, [3 \, m_{qq_{1^+}}^2 m_\Delta^2 + (K\cdot P)^2]\right.\\
&& \rule{15em}{0ex}\left. +m_\Delta[ 2 m_\Delta \ell\cdot K K\cdot P + 3 M (m_{qq_{1^+}}^2 m_\Delta^2 + (K\cdot P)^2)]\right)
\end{eqnarray}
where
\begin{equation}
\sigma_\Delta(\alpha,M,m_{qq_{1^+}},m_\Delta)=(1-\alpha)\, M^2 + \alpha \, m_{qq_{1^+}} - \alpha (1-\alpha)\, m_\Delta^2.
\end{equation}

We employ a symmetry-preserving regularisation scheme. Hence the shift $\ell \to =\ell^\prime+\alpha P$ is permitted, whereafter $O(4)$-invariance entails $\ell^\prime\cdot P=0$ so that one may set
\begin{equation}
\label{replacementsDelta}
\ell \cdot P \to \alpha P^2\,,\;
K\cdot P = (1-\alpha) P^2\,,\;
\ell\cdot K \to \alpha (1-\alpha) P^2\,,
\end{equation}
and therewith obtain \cite{Roberts:2010hu}
\begin{eqnarray}
1 &=& 8 \frac{g_\Delta^2}{M } \frac{E_{qq_{1^+}}^2}{m_{qq_{1^+}}^2}
\int\frac{d^4\ell^\prime}{(2\pi)^4} \int_0^1 d\alpha\,
\frac{(m_{qq_{1^+}}^2 + (1-\alpha)^2 m_\Delta^2)(\alpha m_\Delta + M)}
{[\ell^{^\prime 2} + \sigma_\Delta(\alpha,M,m_{qq_{1^+}},m_\Delta)]^2}\\
&=& \frac{g_\Delta^2}{M}\frac{E_{qq_{1^+}}^2}{m_{qq_{1^+}}^2}\frac{1}{2\pi^2}
\int_0^1 d\alpha\, (m_{qq_{1^+}}^2 + (1-\alpha)^2 m_\Delta^2)(\alpha m_\Delta + M)\overline{\cal C}^{\rm iu}_1(\sigma_\Delta(\alpha,M,m_{qq_{1^+}},m_\Delta))\,.
\end{eqnarray}
This is an eigenvalue problem whose solution yields the mass for the dressed-quark-core of the $\Delta$-resonance.  If one sets $g_\Delta=1$, then $m_\Delta = 1.60\,$GeV.

Construction of the explicit form for the nucleon's Faddeev equation is a straightforward generalisation of the procedure used above for the $\Delta$.  However, it is algebraically more complicated and in App.\,\ref{app:FEnucleon} we simply present the result.  If one sets $g_N=1$, then $m_N=1.27\,$GeV.  It is noteworthy that with $F_{qq_{0^+}}\equiv 0$, one obtains $m_N = 1.14$.  This comparison shows that the scalar-diquark's vector-like component produces approximately $130\,$MeV of repulsion within the nucleon.

One may now read that in the truncation we've described thus far
\begin{equation}
\label{eqmDeltamNdiff}
\delta m := (m_\Delta -m_N) = 0.33\,\mbox{GeV} \; \mbox{cf.} \; \delta_{qq_{10}}:=(m_{qq_{1^+}}-m_{qq_{0^+}}) = 0.28\,\mbox{GeV}.
\end{equation}
We note that these mass differences are correlated; and both vanish together in the limit of infinitely heavy current-quark masses, approaching zero from above as the current-quark mass increases \cite{Cloet:2008fw}.  The latter results are a model-independent consequence of heavy-quark symmetry, kindred to the behaviour of the mass-splitting between vector and pseudoscalar mesons \cite{Bhagwat:2006xi}.

\begin{figure}[t] 
\rightline{\includegraphics[clip,width=0.48\textwidth]{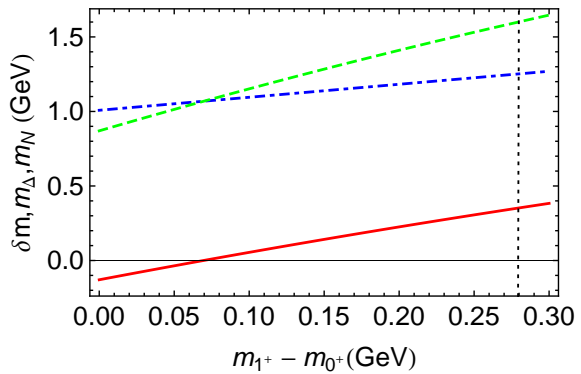}}
\vspace*{-0.25\textwidth}

\parbox{0.47\textwidth}
{\caption{\label{mDeltamNdiff} \emph{Solid curve}, $\delta m:=(m_\Delta - m_N)$; \emph{dashed curve}, $m_\Delta$; \emph{dot-dashed curve}, $m_N$ -- all plotted as a function of $\delta_{qq_{10}}:=(m_{qq_{1^+}}-m_{qq_{0^+}})$.  The vertical dotted line marks the model's predicted values when $g_N=1=g_\Delta$.
}}\vspace*{0.07\textheight}
\end{figure}

The causal connection between $\delta m$ and $\delta_{qq_{10}}$ is readily illustrated. For example, with all other elements held fixed, the latter determines the former.  This is made plain in Fig.\,\ref{mDeltamNdiff}, which depicts the mass-difference $\delta m=(m_\Delta - m_N)$ and the masses $m_\Delta$, $m_N$, all as a function of $\delta_{qq_{10}}$, which was reduced from its model-preferred value by increasing the coupling in the BSE for the axial-vector diquark whilst keeping all other couplings and masses constant.  The behaviour of each curve is readily understood.
The $\Delta$ is an uncomplicated bound state composed of a dressed-quark and an axial-vector diquark.  Since diquark breakup and reformation mediated by the exchange of a dressed-quark is attractive, then decreasing the mass of the axial-vector diquark increases the amount of attraction in the channel because the attraction operates over longer range.  There is nothing in this channel with which the increased attraction can compete, hence $m_\Delta$ drops rapidly with decreasing $\delta_{qq_{10}}$.
The nucleon is more complicated.  Its kernel expresses interference between quark-exchange in the scalar and axial-vector diquark channels, which provides resistance to change because the scalar-diquark properties are held fixed.  Hence, $m_N$ drops more slowly with decreasing $\delta_{qq_{10}}$.
Therefore the behaviour of the mass difference $\delta m$ is driven primarily by the change in $m_\Delta$.
%

\subsection{Pion loops, $m_N$ and $m_\Delta$}
The results described hitherto suggest that whilst corrections to our truncated DSE kernels may have a material impact on $m_N$ and $m_\Delta$ separately, the modification of each is approximately the same, so that the mass-difference, $\delta m$, is largely unaffected by such corrections.  Indeed, this is consistent with an analysis \cite{Young:2002cj} that considers the effect of pion loops, which are explicitly excluded in the rainbow-ladder truncation \cite{Eichmann:2008ae}: whilst the individual masses are reduced by roughly $300\,$MeV, the mass difference, $\delta m$, increases by only $50\,$MeV.

We emphasise that it is essential not to miscount when incorporating the effect of pseudoscalar meson loops.  In practical calculations these effects divide into two distinct types.  The first is within the gap equation, where pseudoscalar meson loop corrections to the dressed-quark-gluon vertex act to reduce uniformly the mass-function of a dressed-quark \cite{Eichmann:2008ae,Cloet:2008fw}.  This effect can be pictured as a single quark emitting and reabsorbing a pseudoscalar meson.  It can be mocked-up by simply choosing the parameters in the gap equation's kernel so as to obtain a dressed-quark mass-function that is characterised by a mass-scale of approximately $400\,$MeV.  Such an approach has implicitly been widely employed with phenomenological success \cite{Roberts:2000aa,Maris:2003vk,Roberts:2007jh}, and is that which we employ herein.

The second type of correction arises in connection with bound-states and may be likened to adding pseudoscalar meson exchange \emph{between} dressed-quarks within the bound-state \cite{Hollenberg:1992nj,Alkofer:1993gu,Ishii:1998tw,Pichowsky:1999mu}, as opposed to the first type of effect; i.e., emission and absorption of a meson by the same quark.  The type-2 contribution is that computed in typical evaluations of meson-loop corrections to hadron observables based on a point-hadron Lagrangian.  This fact should be borne in mind when using formulae, such as those in Ref.\,\cite{Young:2002cj}, to estimate the size of meson-loop corrections to bound-state masses computed using the DSEs.

This discussion establishes that it is correct to use such formulae herein, just as it was in Refs.\,\cite{Hecht:2002ej,Cloet:2008re}.  Their straightforward application using a common meson-baryon form-factor mass-scale of $0.8\,$GeV yields a shift of $(-300\,$MeV$)$ in $m_N$ and $(-270\,$MeV$)$ in $m_\Delta$, from which one may infer that our type-2 uncorrected Faddeev equations should produce $m_N=1.24\,$GeV and $m_\Delta=1.50\,$GeV, values which are plainly of the appropriate size.  For the $\Delta$-resonance there is another estimate, which is arguably more sophisticated.  Namely, that produced by the Excited Baryon Analysis Center (EBAC) \cite{Suzuki:2009nj}, which used a realistic coupled-channels model to remove meson dressing from the $\Delta$ and expose a dressed-quark-core bare-mass of $1.39\,$GeV.  Following these observations we return to Eq.\,(\ref{staticexchange}) and choose
\begin{equation}
\label{gNgDelta}
g_N= 1.18 \,,\; g_\Delta = 1.56\,\; \Rightarrow m_N=1.14\,\mbox{GeV}, m_\Delta = 1.39\,\mbox{GeV}, \delta m = 0.25\,\mbox{GeV}
\end{equation}
because the listed outcomes of this choice are consistent with the information presented above and Refs.\,\cite{Ishii:1998tw,Hecht:2002ej,Cloet:2008re}.

\begin{figure}[t] 
\includegraphics[clip,width=0.48\textwidth]{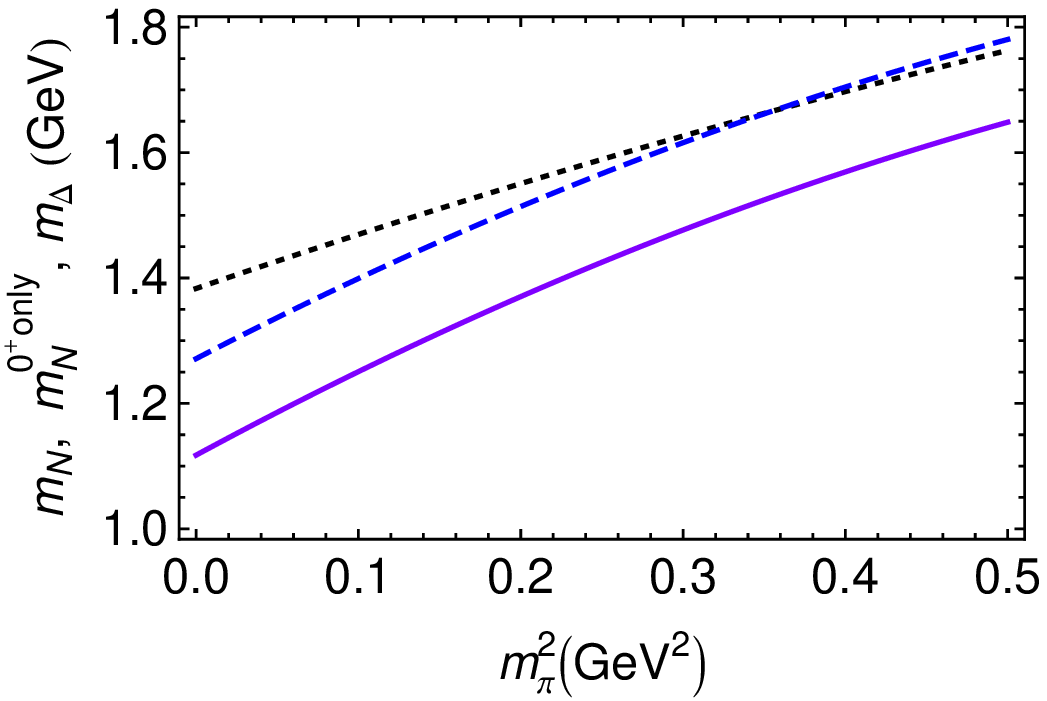}\hspace*{1em}
\includegraphics[clip,width=0.48\textwidth]{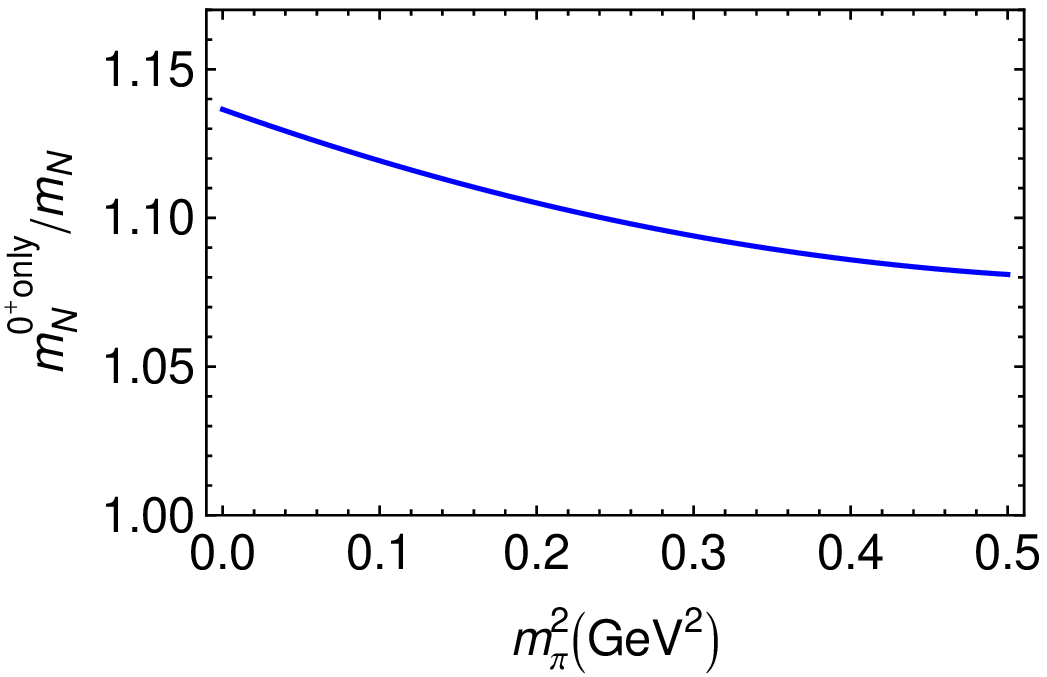}
\caption{\label{nucleonO0}
\emph{Left panel}  -- Evolution with current-quark mass of the: nucleon mass, $m_N$ (\emph{solid curve}); scalar-diquark-only nucleon mass, $m_N^{0^+\,\mbox{only}}$ (dashed curve); and $\Delta$ mass, $m_\Delta$.
\emph{Right panel} -- Ratio $m_N^{0^+\,\mbox{only}}/m_N$ as a function of current-quark mass.  It drops by just 5\% on the domain depicted.
In all panels here and below the current-quark mass is expressed through the computed value of $m_\pi^2$: $m_\pi^2=0.49\,$GeV$^2$ marks the $s$-quark current-mass.
}
\end{figure}

\subsection{Evolution of ground-state masses with current-quark mass}
Within a framework such as we employ it is straightforward to map the evolution with increasing current-quark mass, $m$, of bound-state masses and the splittings between them.  In rainbow-ladder truncation it is only the current-mass which varies because the dressed-quark-gluon vertex is independent of $m$.   Notably, in systems related to the pseudoscalar- and vector-meson channels, non-resonant corrections to the rainbow-ladder truncation do not materially affect mass-splittings \cite{Bhagwat:2004hn}.  Hence the interpretation of our results on splittings between the bound-states' quark-cores should be robust.

The nucleon is constituted from scalar and axial-vector diquark correlations.  It is therefore of interest to determine the effect of the axial-vector correlation on the nucleon's mass.  This can be read from Fig.\,\ref{nucleonO0}: the axial-vector diquark-correlation provides attraction in the nucleon channel, in an amount which is almost independent of current-quark mass up to values matching the $s$-quark current-mass.  This attraction ensures the nucleon remains lighter than the $\Delta$ for all values of current-quark mass.  A nucleon constituted solely from a dressed-quark and scalar-diquark correlation will finally become more massive than the $\Delta$.  In our case, this occurs for $m\gsim 0.8\,m_s$.

\begin{figure}[t] 
\includegraphics[clip,width=0.48\textwidth]{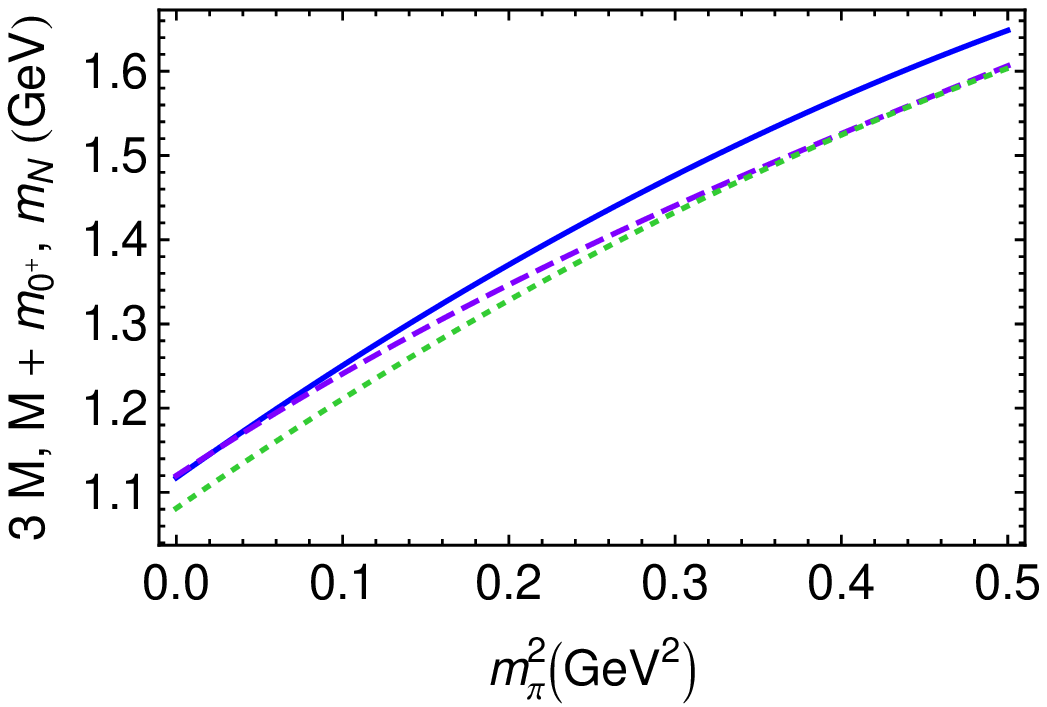}\hspace*{1em}
\includegraphics[clip,width=0.48\textwidth]{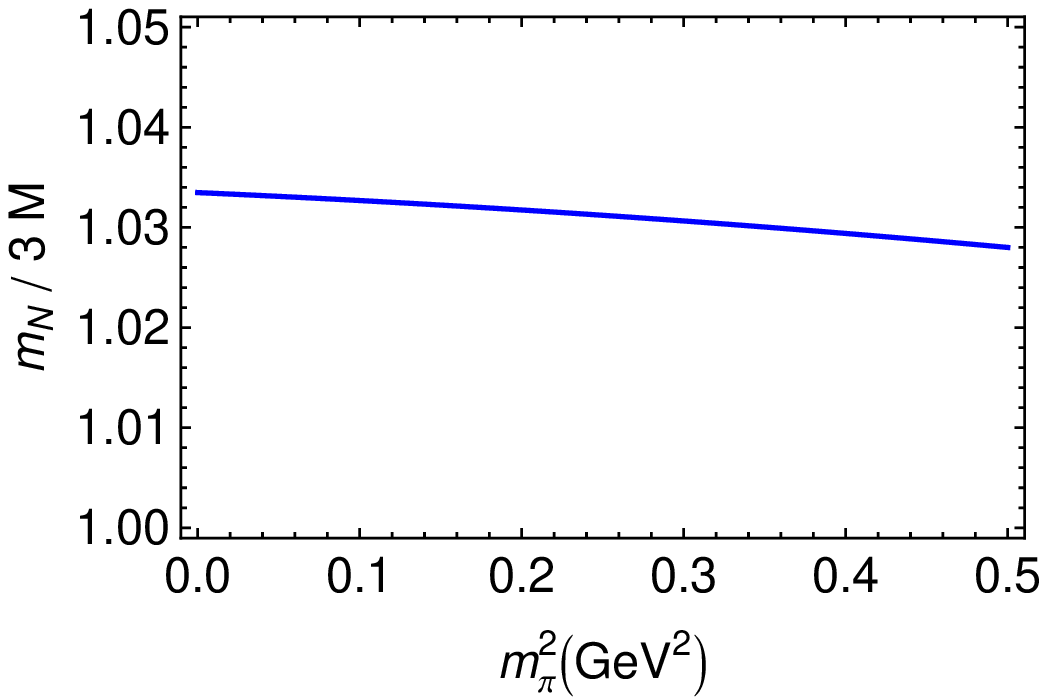}
\caption{\label{nucleonMm0}
\emph{Left panel}  -- Evolution with current-quark mass of the: nucleon mass, $m_N$ (\emph{solid curve}); the sum $[M+m_{qq_{0^+}}]$ (\emph{dashed curve}); and $ 3 \, M$ (\emph{dotted curve}).
\emph{Right panel} -- Evolution with current-quark mass of the ratio $m_N/[3 M]$, which varies by less-than 1\% on the domain depicted.
%
}
\end{figure}

In the left panel of Fig.\,\ref{nucleonMm0} we depict the evolution with current-quark mass of $m_N$, the mass of the nucleon's dressed-quark-core, and compare it with the evolution of the combinations $[M+m_{qq_{0^+}}]$ and $3 M$.  In conjunction with the ratio in the right panel, it is evident that $m_N$ is given by $3 M$ \emph{plus} a small contribution that is almost independent of current-quark mass.\footnote{In this near-proportionality there is a similarity with models of the constituent-quark type.  NB. It is our confining regularisation of the contact-interaction which enables the mass of the bound-state to be greater than that contained in the masses of its constituents.}  Hence, if one inflates $M$ in an attempt to anticipate type-1 pseudoscalar-meson vertex-corrections within the gap equation \cite{Eichmann:2008ae}, then the nucleon mass will increase commensurately.  It is noteworthy that for $m\gsim m_s$, $m_{qq_{0^+}} \approx 2 M$.

\begin{figure}[t] 
\includegraphics[clip,width=0.48\textwidth]{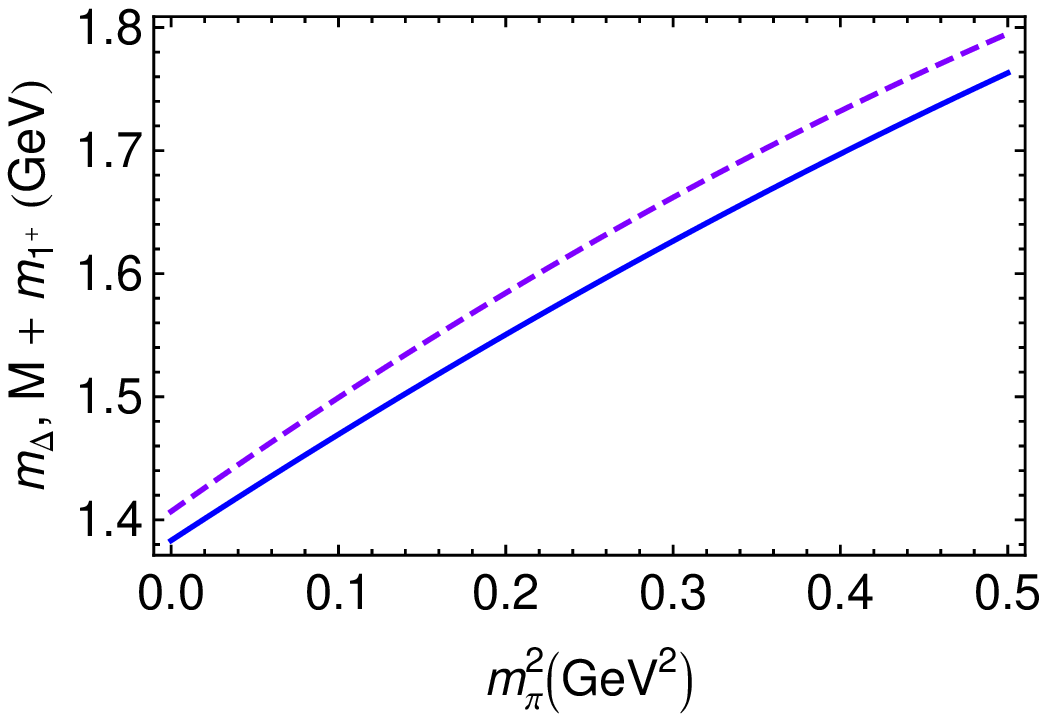}\hspace*{1em}
\includegraphics[clip,width=0.48\textwidth]{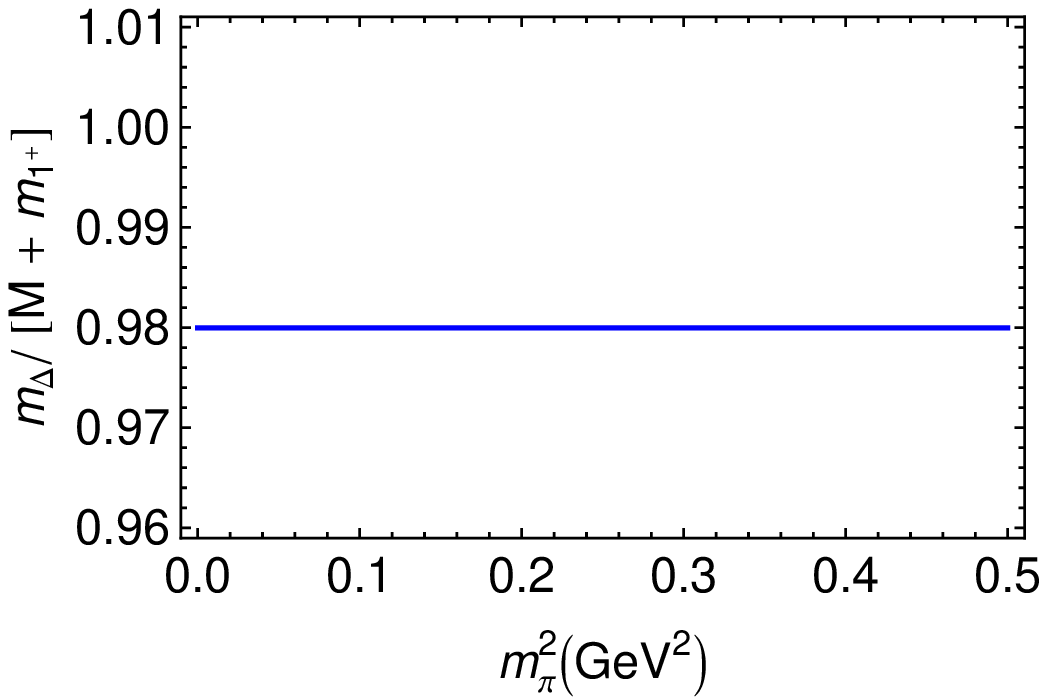}
\caption{\label{Fig4}
\emph{Left panel}  -- Evolution with current-quark mass of the: $\Delta$ mass, $m_\Delta$ (\emph{solid curve}); and $[M+m_{qq_{1^+}}]$ (dashed curve).
\emph{Right panel} -- Evolution with current-quark mass of the ratio $m_\Delta/[M + m_{qq_{1^+}}]$, which does not vary noticeably on the domain depicted.
%
}
\end{figure}

In the left panel of Fig.\,\ref{Fig4} we display the evolution with current-quark mass of $m_\Delta$, the dressed-quark-core mass of the $\Delta$, along with that of the sum $M+m_{qq_{1^+}}$.  As the right panel makes plain, to a very good level of approximation $m_\Delta = M+m_{qq_{1^+}}$.
Recall now that $m_{qq_{1^+}}>m_{qq_{0^+}}$ for all values of current-quark mass, with $[m_{qq_{1^+}}-m_{qq_{0^+}}]\to 0^+$ as $m\to \infty$.  It follows that if one inflates $M$ in an attempt to anticipate type-1 pseudoscalar-meson vertex-corrections within the gap equation \cite{Eichmann:2008ae}, then the $\Delta$ mass will increase by an amount far larger than is seen with the nucleon.  Indeed, the amount will be commensurate with the inflation in $3 M$ \emph{plus} the scaled increase of the bound-state energy-excess $[m_{qq_{1^+}}-m_{qq_{0^+}}]$.  This must be understood if erroneous conclusions about the nature of the quark-core of the $\Delta$ are to be avoided.

\begin{figure}[t]
\includegraphics[clip,width=0.48\textwidth]{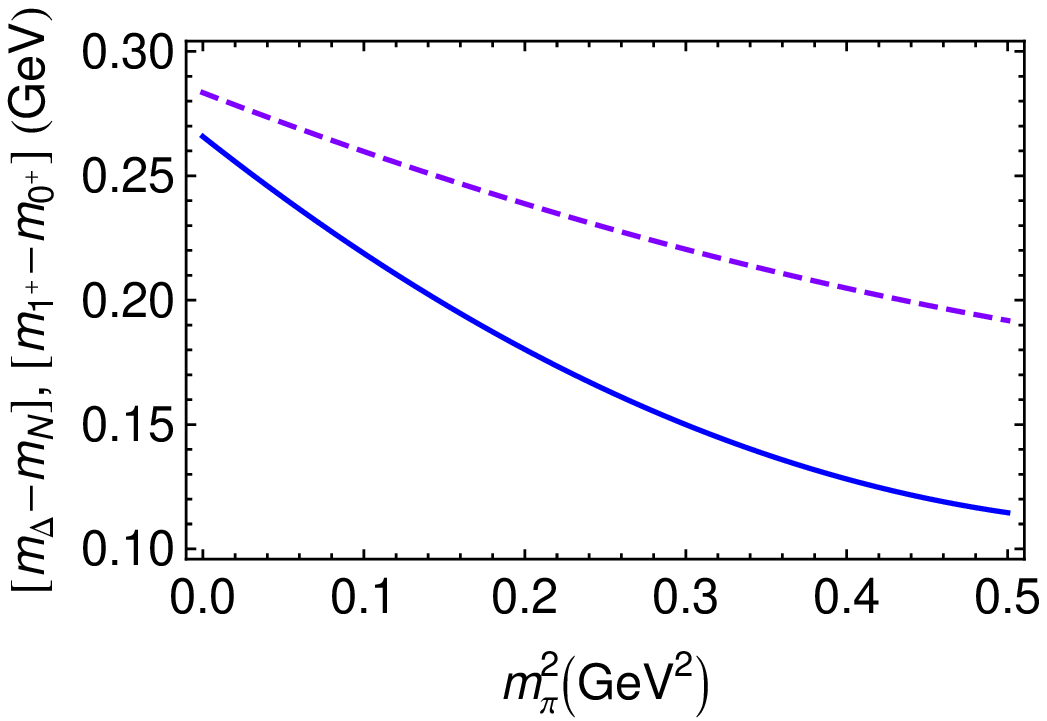}\hspace*{1em}
\includegraphics[clip,width=0.48\textwidth]{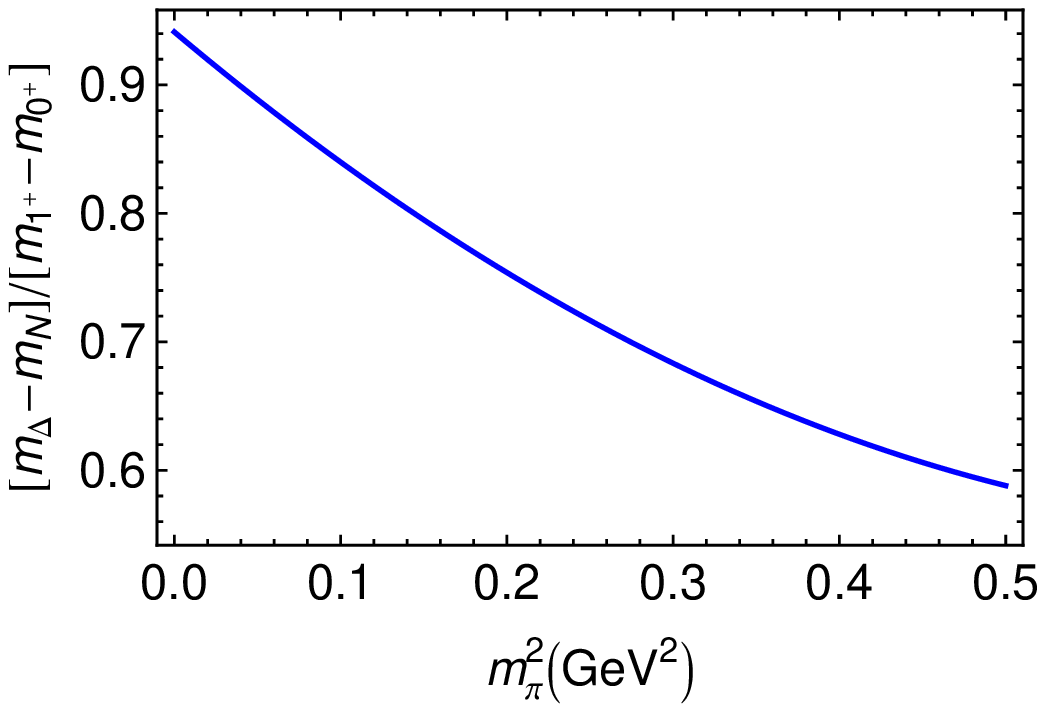}
\caption{\label{Fig5}
\emph{Left panel}  -- Evolution with current-quark mass of the: $\Delta$-nucleon mass difference $[m_\Delta - m_N]$ (\emph{solid curve}); and axial-vector--scalar-diquark mass difference $[m_{qq_{1^+}} - m_{qq_{0^+}}]$ (\emph{dashed curve}).
\emph{Right panel} -- Evolution with current-quark mass of the ratio $[m_\Delta - m_N]/[m_{qq_{1^+}} - m_{qq_{0^+}}]$, which falls by 40\% on the domain depicted.
%
}
\end{figure}

Figure~\ref{Fig5} depicts the evolution with current-quark mass of the mass differences $\delta m = [m_\Delta - m_N]$ and $\delta_{qq_{11}} = [m_{qq_{1^+}} - m_{qq_{0^+}}]$.  As was seen elsewhere \cite{Cloet:2008fw}, with increasing $m$, $\delta m$ decreases more rapidly than $\delta_{qq_{11}}$ and both mass-differences approach zero uniformly from above.  The difference between the rates of decrease is readily explained.  The difference $\delta_{qq_{11}}$ becomes smaller because increasing the quark mass suppresses hyperfine interactions, as demonstrated elsewhere \cite{Bhagwat:2006xi} for the analogous case of the vector--pseudoscalar-meson mass difference.  This reduction introduces circumstances analogous to those illustrated in Fig.\,\ref{mDeltamNdiff}, which may now be read as follows.  The nucleon is lighter than the $\Delta$ owing to attraction provided by quark-exchange originating from both scalar- and axial-vector-diquark correlations.  With increasing quark mass, not only do the constituents of both systems approach common masses but the additional attraction is diminished.  Hence $m_N$ increases more rapidly than does the mass of the $\Delta$, whose simpler structure means it draws attraction from only one source in the Faddeev equation.

\begin{figure}[t]
\rightline{\includegraphics[clip,width=0.48\textwidth]{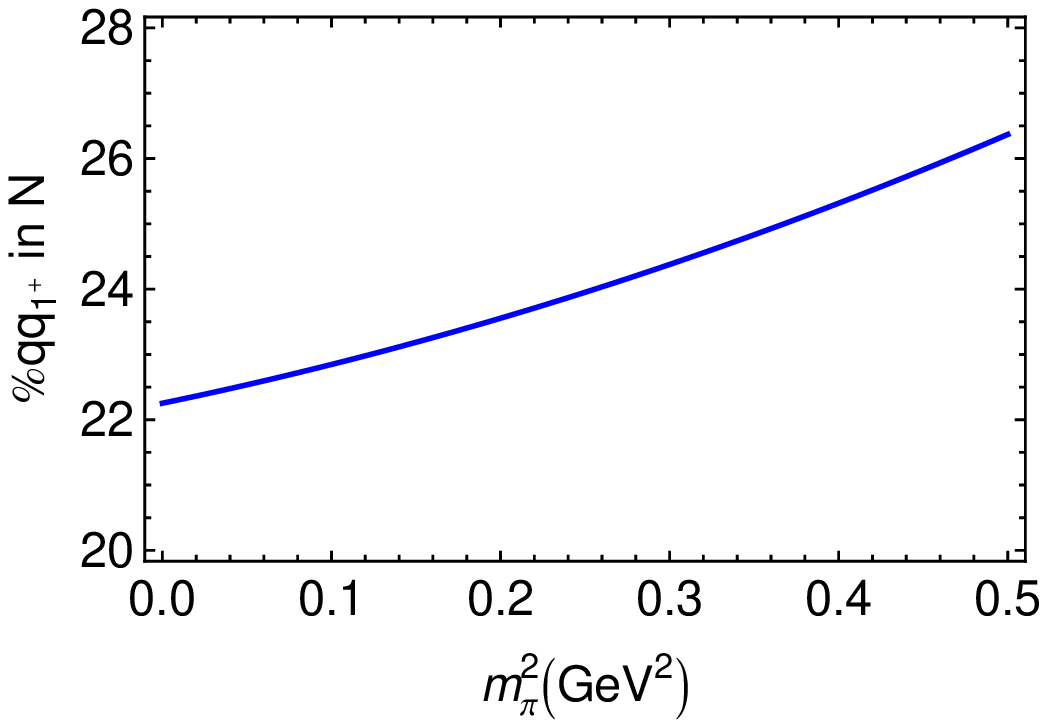}}
\vspace*{-0.20\textheight}

\parbox{0.47\textwidth}{
\caption{\label{Fig6}
Evolution with current-quark mass of the contribution made by axial-vector-diquark correlations to the unit-normalisation of the nucleon's Faddeev amplitude.  At the model-preferred current-quark mass, the probability is 22\%.  This rises to 26\% at $m=m_s$.
%
}}\vspace*{0.08\textheight}
\end{figure}

There is another interesting consequence of increasing current-quark mass, which concerns the axial-vector-diquark content of the nucleon.  As $\delta_{qq_{11}}$ becomes smaller, there is less to distinguish between the range and nature of the attractive interactions provided by the scalar- and axial-vector-diquark correlations.  Hence, as shown in Fig.\,\ref{Fig6}, the relative strength of the axial-vector correlation increases with $m$.
This relative strength has a material impact on nucleon properties.  For example, in the electromagnetic form factor calculations described in Ref.\,\cite{Cloet:2008re}, the photon-nucleon interaction involves an axial-vector diquark correlation with 40\% probability.  This value: is crucial in fixing the location of a zero in the ratio $F_1^{p,d}(Q^2)/F_1^{p,u}(Q^2)$ \cite{Roberts:2010hu}; determines the $x=1$ value for ratio of nucleon structure functions $F_2^n/F_2^p = 0.36$ \cite{Holt:2010vj}; and entails that in the nucleon's rest frame just 37\% of the total spin of the nucleon is contained within components of the Faddeev amplitude which possess zero quark orbital angular momentum \cite{Cloet:2007pi}.

\subsection{Radial excitations and parity partners}
In analogy with mesons, the leading Tchebychev moment of the bound-state amplitude for a baryon's first radial excitation should possess a single zero.  Whilst the truncation we employ cannot generate such a zero, it is possible to estimate masses for these states by employing the expedient described in Sect.\,\ref{mesonradial}.  With the zero located as prescribed in Eq.\,(\ref{zerovalue}), no new parameters are introduced.  To illustrate, the mass of the first radial excitation of the $\Delta$, $m_{\Delta^\ast}$, is determined via
\begin{eqnarray}
1 &=& \frac{g_\Delta^2}{M}\frac{E_{qq_{1^+}}^2}{m_{qq_{1^+}}^2}\frac{1}{2\pi^2}
\int_0^1 d\alpha\, (m_{qq_{1^+}^2} + (1-\alpha)^2 m_{\Delta^\ast}^2)(\alpha \, m_{\Delta^\ast} + M)\overline{\cal F}^{\rm iu}_1(\sigma_\Delta(\alpha,M,m_{qq_{1^+}},m_{\Delta^\ast}))\,.
\end{eqnarray}

In a more general setting one might imagine that a baryon's first radial excitation could be an admixture of two components: one with a zero in the Faddeev amplitude, describing a radial excitation of the quark-diquark system; and the other with a zero in the diquark's Bethe-Salpeter amplitude, which represents an internal excitation of the diquark.  The procedure in Sect.\,\ref{mesonradial} can conceivably distinguish between these components via a mixing term whose strength is $\propto E_{qq_{1^+}}E_{qq^\ast_{1^+}}$.  Owing to orthogonality of the two-body ground- and first-radially-excited states, we anticipate that this mixing term is negligible.  Under this assumption, a baryon's first radial excitation is predominantly a radial excitation of the quark-diquark system.  Should a state constituted from a radially-excited diquark exist, then its mass will be larger because $[E_{qq^\ast_{1^+}}^2/m_{qq^\ast_{1^+}}^2]<[E_{qq_{1^+}}^2/m_{qq_{1^+}}^2]$.

Given the preceding discussion, it will not be surprising that we define bound-state equations for the parity-partners of the ground- and first-radially-excited-states of the nucleon and $\Delta$-resonance by making the replacements
\begin{equation}
E_{qq_{J^P}} \to E_{qq_{J^{-P}}}\,, \; m_{qq_{J^P}} \to m_{qq_{J^{-P}}}
\end{equation}
in the appropriate Faddeev equations.  For example, we determine the mass of the $J^P = \frac{3}{2}^{-}$ $\Delta$-state from
\begin{equation}
1 = \frac{g_\Delta^2}{M}\frac{E_{qq_{1^-}}^2}{m_{qq_{1^-}}^2}\frac{1}{2\pi^2}
\int_0^1 d\alpha\, (m_{qq_{1^-}^2} + (1-\alpha)^2 m_{\Delta\frac{3}{2}^-}^2)(\alpha \, m_{\Delta\frac{3}{2}^-} + M)\overline{\cal C}^{\rm iu}_1(\sigma_\Delta(\alpha,M,m_{qq_{1^-}},m_{\Delta\frac{3}{2}^-}))\,.
\end{equation}

\subsection{Computed baryon spectrum}
In Table~\ref{Table:baryon} we list our computed results for the dressed-quark-core masses of the nucleon and $\Delta$, their first radial excitations (denoted by ``$\ast$''), and the parity-partners of these states.  These masses cannot be compared directly with experiment because the kernels employed in their calculation do not incorporate the effect of meson loops.  However, a fair comparison may be made with bare-masses inferred from sophisticated coupled-channels analyses of $\pi N$ scattering data up to $W\lsim 2\,$GeV \cite{Suzuki:2009nj,Gasparyan:2003fp}.

\begin{table}[t]
\caption{
\emph{Row-1}: Dressed-quark-core masses for nucleon and $\Delta$, their first radial excitations (denoted by ``$\ast$''), and the parity-partners of these states, computed with $g_N=1.18$, $g_\Delta=1.56$, and the parameter values in Eq.\,(\protect\ref{gSO}) and Table~\protect\ref{Table:static}.  The errors on the masses of the radial excitations indicate the effect of shifting the location of the zero according to Eq.\,(\protect\ref{zerovalue}).
\emph{Row-2}: Bare-masses inferred from a coupled-channels analysis at the Excited Baryon Analysis Center (EBAC) \protect\cite{Suzuki:2009nj}.  EBAC's method does not provide a bare nucleon mass.
\emph{Row-3}: Bare masses inferred from the coupled-channels analysis described in Ref.\,\protect\cite{Gasparyan:2003fp}, which describes the Roper resonance as dynamically-generated.
In both these rows, ``{\ldots}'' indicates states not found in the analysis.
A visual comparison of these results is presented in Fig.\,\protect\ref{Fig7}.
\label{Table:baryon}
}
\begin{center}
\begin{tabular*}
{\hsize}
{
l@{\extracolsep{0ptplus1fil}}
|c@{\extracolsep{0ptplus1fil}}
c@{\extracolsep{0ptplus1fil}}
c@{\extracolsep{0ptplus1fil}}
c@{\extracolsep{0ptplus1fil}}
|c@{\extracolsep{0ptplus1fil}}
c@{\extracolsep{0ptplus1fil}}
c@{\extracolsep{0ptplus1fil}}
c@{\extracolsep{0ptplus1fil}}}\hline

& $m_N$ & $m_{N^\ast}$ & $m_{N \frac{1}{2}^{-}}$ & $m_{N^\ast \frac{1}{2}^{-}}$
& $m_\Delta$ & $m_{\Delta^\ast}$ & $m_{\Delta \frac{3}{2}^{-}}$ & $m_{\Delta^\ast \frac{3}{2}^{-}}$\rule{0em}{2.5ex}\\
PDG label~ & $N$  & $N(1440)\,P_{11}$ & $N(1535)\,S_{11}$ & $N(1650)\,S_{11}$ & $\Delta(1232)\,P_{33}$ & $\Delta(1600)\,P_{33}$ & $\Delta(1700)\,D_{33}$ & $\Delta(1940)\,D_{33}$\rule{0em}{2.5ex}\\\hline
%
This work~ & 1.14 & 1.82$\pm0.07$ & 2.22  & $2.29\pm0.02$ & 1.39 & $1.85\pm 0.05$ & 2.25 & $2.33\pm 0.02$\rule{0em}{2.5ex}\\
EBAC~      &      & 1.76 & 1.80  & 1.88 & 1.39 & \ldots & 1.98 & \ldots\rule{0em}{2.5ex}\\
J\"ulich~  & 1.24 & none & 2.05  & 1.92 & 1.46 & \ldots & 2.25 & \ldots\rule{0em}{2.5ex}\\\hline
\end{tabular*}
\end{center}
\end{table}

The predictions of our model for the baryon's dressed-quark-core match the bare-masses determined in Ref.\,\cite{Gasparyan:2003fp} with a root-mean-square (rms) relative-error of 10\%.  Notably, however, we find a quark-core to the Roper resonance, whereas within the J\"ulich coupled-channels model this structure in the $P_{11}$ partial wave is unconnected with a bare three-quark state.  In connection with EBAC's analysis, our predictions for the bare-masses agree within a rms relative-error of 14\%.  Notably, EBAC does find a dressed-quark-core for the Roper resonance, at a mass which agrees with our prediction.

We also predict dressed-quark-core states associated with radial excitations of the $\Delta$-resonance.  Allowing for a reduction by $\lsim 160\,$MeV expected from coupled channels effects, our estimate for the mass of the three-star $\Delta(1600)\,P_{33}$-resonance is in agreement with contemporary experiment \cite{Nakamura:2010zzi}.  The same is true of our result for this state's parity-partner, $\Delta(1940)\,D_{33}$.

In this connection we observe that the EBAC analysis does not find any sign of the $\Delta(1600)\,P_{33}$ or $\Delta(1940)\,D_{33}$ resonances.  This is plausibly an indication of a limitation in the method employed to complete the difficult task of reaching into the complex plane in order to locate the poles associated with these resonances.

It is also worth remarking that the J\"ulich analysis of the $I=\frac{3}{2}$-channel has been revisited \cite{Doring:2010ap}, with new bare masses being reported: 1535\,MeV for the $\Delta(1232)$ and 3442\,MeV for $\Delta(1700)$.  However, this study finds that the bare-mass values depend sensitively upon precisely which channel-couplings are included in the model.  It is notable that this analysis identifies the $\Delta(1600)\,P_{33}$ with a broad, dynamically-generated resonance.

\begin{figure}[t]
\rightline{\includegraphics[clip,width=0.60\textwidth]{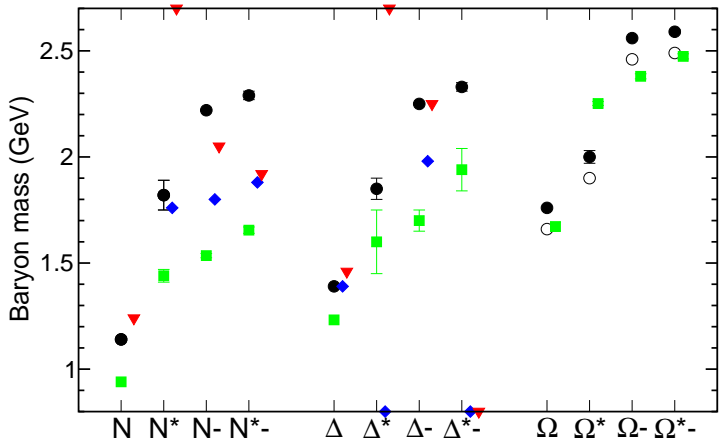}}
\vspace*{-0.255\textheight}

\parbox{0.38\textwidth}{
\caption{\label{Fig7}
Visual comparison of our computed baryon masses (\emph{filled circles}) with bare masses determined in Ref.\,\protect\cite{Suzuki:2009nj} (\emph{filled diamonds}) and Ref.\,\protect\cite{Gasparyan:2003fp} (\emph{filled triangles}).
For the coupled-channels models a symbol at the lower extremity indicates that no associated state is found in the analysis, whilst a symbol at the upper extremity indicates that the analysis reports a dynamically-generated resonance with no corresponding bare-baryon state.
In connection with the $\Omega$-baryons the \emph{open-circles} represent our results shifted downwards by $100\,$MeV.  [See discussion after Eq.\,(\protect\ref{Omegabaryons}).]
The \emph{filled-squares} report masses tabulated in Ref.\,\protect\cite{Nakamura:2010zzi}.
}}\vspace*{0.02\textheight}
\end{figure}

With the information now available we can also report dressed-quark-core masses for the decuplet $\Omega^-$ baryon; viz.,
\begin{equation}
\label{Omegabaryons}
\begin{array}{cccc}
m_\Omega & m_{\Omega^\ast} & m_{\Omega \frac{3}{2}^{-}} & m_{\Omega^\ast\frac{3}{2}^{-}}\rule{0em}{2.5ex}\\
%
1.76 & ~~ 2.00 \pm 0.03 ~~ & 2.56 & ~~ 2.59 \pm 0.01 ~~\rule{0em}{2.5ex} \\
%
\end{array}\,.
\end{equation}
Only four $\Omega$-baryons are listed in Ref.\,\cite{Nakamura:2010zzi}: $\Omega^-(1670)$, a four-star $J^P=\frac{3}{2}^+$ state; $\Omega^-(2250)$, a three-star state; and $\Omega^-(2380)$, $\Omega^-(2470)$, two two-star states.  The spin-parity of the last three resonances is currently unknown.  To place our computed $\Omega$-masses in context we observe that at $m=m_s$ the computed vector-meson dressed-quark-core mass is $m_\phi=1.13$, which is $110\,$MeV above the experimental value.  Notably, pseudoscalar-meson loop corrections are estimated to reduce the core mass by $\simeq 100\,$MeV \cite{Eichmann:2008ae,Leinweber:2001ac}.  Furthermore, a similar analysis indicates that, at $m_\pi^2=0.5\,$GeV$^2$, pseudoscalar-meson loop corrections in $\Delta$-like systems produce a ($-100\,$MeV) shift in the mass of the baryon's dressed-quark core \cite{Young:2002cj}.  A comparison with experiment is presented in Fig.\,\ref{Fig7}.

\section{Epilogue}
\label{epilogue}
We presented the first DSE-based calculation of the light hadron spectrum that simultaneously correlates the dressed-quark-core masses of meson and baryon ground- and excited-states within a single symmetry-preserving framework.  Isospin symmetry was assumed, with $m_u=m_d=m=7\,$MeV producing a physical pion mass; and five parameters were used to define the gap-, Bethe-Salpeter- and Faddeev-equations.  In a comparison with relevant quantities, we recorded a value of 13\% for the overall root-mean-square-relative-error$/$degree-of freedom ($\overline{\mbox{rms}}$).  Notable amongst our results is agreement between the computed masses for baryon dressed-quark-cores and the bare masses employed in modern dynamical coupled-channels models of pion-nucleon reactions.

In connection with mesons we capitalised on recent progress in understanding the far-reaching effects of dynamical chiral symmetry breaking within the Bethe-Salpeter kernel to improve upon the rainbow-ladder truncation in the scalar and axial-vector channels.  This enabled us to obtain $\overline{\mbox{rms}}_{\bar q q}= 13\%$, a feature which also has a significant collateral impact on the baryon spectrum owing to the connection between the Bethe-Salpeter equations for mesons and diquark-correlations.

In comparison with relevant quantities our predicted baryon masses yield $\overline{\mbox{rms}}_{qqq}=14$\%.  Furthermore, our analysis provides insight into numerous aspects of baryon structure.  For example, we explained that in practical formulations of the baryon bound-state problem, there are two distinct types of pseudoscalar-meson-loop correction: one intrinsic to the gap equation; and another restricted to bound-state kernels.  It is the latter which is expressed in the formulae typically used to estimate meson-cloud contributions to baryon masses.

We also demonstrated relationships between the masses of the nucleon and $\Delta$  ($m_N$, $m_\Delta$), and those of the dressed-quark and diquark correlations they contain (M, $m_{qq_{0^+}}$, $m_{qq_{1^+}}$).  For example, we established a causal connection between $[m_\Delta - m_N]$ and $[m_{qq_{1^+}}-m_{qq_{0^+}}]$; demonstrated that $m_N \approx 3 M$ and $m_\Delta \approx M+m_{qq_{1^+}}$; illustrated the simplicity of the $\Delta$'s internal structure and the consequences of this; and showed that the relative strength of the axial-vector diquark-correlation within the nucleon grows with current-quark mass.

At the core of our analysis is a symmetry-preserving treatment of a vector-vector contact interaction.  Our body of results confirms that this is a useful tool for the study of phenomena characterised by probe momenta less-than the dressed-quark mass.  It is now important to use this foundation in the computation of baryon elastic and transition form factors.  That will provide information which is crucial in using experimental data on such observables as a tool for charting the nature of the quark-quark interaction at long-range \cite{Aznauryan:2009da}.

\begin{acknowledgements}
We acknowledge valuable discussions with A.~Bashir, M.~D\"oring, S.~Krewald, T.\,S-H.~Lee, C.~Hanhart and S.\,M.~Schmidt.
This work was supported by:
Forschungszentrum J\"ulich GmbH;
the U.\,S.\ Department of Energy, Office of Nuclear Physics, contract nos.~DE-FG03-97ER4014 and DE-AC02-06CH11357;
and the Department of Energy's Science Undergraduate Laboratory Internship programme.
\end{acknowledgements}

\appendix
\setcounter{equation}{0}
\renewcommand{\theequation}{\Alph{section}.\arabic{equation}}
\section{Euclidean Conventions}
\label{App:EM}
In our Euclidean formulation:
\begin{equation}
p\cdot q=\sum_{i=1}^4 p_i q_i\,;
\end{equation}
\begin{equation}
\{\gamma_\mu,\gamma_\nu\}=2\,\delta_{\mu\nu}\,;\;
\gamma_\mu^\dagger = \gamma_\mu\,;\;
\sigma_{\mu\nu}= \frac{i}{2}[\gamma_\mu,\gamma_\nu]\,; \;
{\rm tr}\,[\gamma_5\gamma_\mu\gamma_\nu\gamma_\rho\gamma_\sigma]=
-4\,\epsilon_{\mu\nu\rho\sigma}\,, \epsilon_{1234}= 1\,.
\end{equation}

A positive energy spinor satisfies
\begin{equation}
\bar u(P,s)\, (i \gamma\cdot P + M) = 0 = (i\gamma\cdot P + M)\, u(P,s)\,,
\end{equation}
where $s=\pm$ is the spin label.  It is normalised:
\begin{equation}
\bar u(P,s) \, u(P,s) = 2 M \,,
\end{equation}
and may be expressed explicitly:
\begin{equation}
u(P,s) = \sqrt{M- i {\cal E}}
\left(
\begin{array}{l}
\chi_s\\
\displaystyle \frac{\vec{\sigma}\cdot \vec{P}}{M - i {\cal E}} \chi_s
\end{array}
\right)\,,
\end{equation}
with ${\cal E} = i \sqrt{\vec{P}^2 + M^2}$,
\begin{equation}
\chi_+ = \left( \begin{array}{c} 1 \\ 0  \end{array}\right)\,,\;
\chi_- = \left( \begin{array}{c} 0\\ 1  \end{array}\right)\,.
\end{equation}
For the free-particle spinor, $\bar u(P,s)= u(P,s)^\dagger \gamma_4$.

The spinor can be used to construct a positive energy projection operator:
\begin{equation}
\label{Lplus} \Lambda_+(P):= \frac{1}{2 M}\,\sum_{s=\pm} \, u(P,s) \, \bar
u(P,s) = \frac{1}{2M} \left( -i \gamma\cdot P + M\right).
\end{equation}

A negative energy spinor satisfies
\begin{equation}
\bar v(P,s)\,(i\gamma\cdot P - M) = 0 = (i\gamma\cdot P - M) \, v(P,s)\,,
\end{equation}
and possesses properties and satisfies constraints obtained via obvious analogy
with $u(P,s)$.

A charge-conjugated Bethe-Salpeter amplitude is obtained via
\begin{equation}
\label{chargec}
\bar\Gamma(k;P) = C^\dagger \, \Gamma(-k;P)^{\rm T}\,C\,,
\end{equation}
where ``T'' denotes a transposing of all matrix indices and
$C=\gamma_2\gamma_4$ is the charge conjugation matrix, $C^\dagger=-C$.  We note that
\begin{equation}
C^\dagger \gamma_\mu^{\rm T} \, C = - \gamma_\mu\,, \; [C,\gamma_5] = 0\,.
\end{equation}

In describing the $\Delta$ resonance we employ a Rarita-Schwinger spinor to
unambiguously represent a covariant spin-$3/2$ field.  The positive energy
spinor is defined by the following equations:
\begin{equation}
\label{rarita}
(i \gamma\cdot P + M)\, u_\mu(P;r) = 0\,,\;
\gamma_\mu u_\mu(P;r) = 0\,,\;
P_\mu u_\mu(P;r) = 0\,,
\end{equation}
where $r=-3/2,-1/2,1/2,3/2$.  It is normalised:
\begin{equation}
\bar u_{\mu}(P;r^\prime) \, u_\mu(P;r) = 2 M\,,
\end{equation}
and satisfies a completeness relation
\begin{equation}
\label{Deltacomplete}
\frac{1}{2 M}\sum_{r=-3/2}^{3/2} u_\mu(P;r)\,\bar u_\nu(P;r) =
\Lambda_+(P)\,R_{\mu\nu}\,,
\end{equation}
where
\begin{equation}
R_{\mu\nu} = \delta_{\mu\nu} \mbox{\boldmath $I$}_{\rm D} -\frac{1}{3} \gamma_\mu \gamma_\nu +
\frac{2}{3} \hat P_\mu \hat P_\nu \mbox{\boldmath $I$}_{\rm D} - i\frac{1}{3} [ \hat P_\mu
\gamma_\nu - \hat P_\nu \gamma_\mu]\,,
\end{equation}
with $\hat P^2 = -1$, which is very useful in simplifying the positive energy
$\Delta$'s Faddeev equation.

\setcounter{equation}{0}
\section{Bethe-Salpeter equations}
\label{app:groundBSEs}
\subsection{Pseudoscalar mesons and scalar diquarks}
\label{app:groundBSEspion}
The explicit form of Eq.\,(\ref{Gamma-eq}) is:
\begin{equation}
\label{bsefinal0}
\left[
\begin{array}{c}
E_{0^-}(P)\\
F_{0^-}(P)
\end{array}
\right]
= \frac{1}{3\pi^2 m_G^2}
\left[
\begin{array}{cc}
{\cal K}_{EE}^\pi & {\cal K}_{EF}^\pi \\
{\cal K}_{FE}^\pi & {\cal K}_{FF}^\pi
\end{array}\right]
\left[\begin{array}{c}
E_{0^-}(P)\\
F_{0^-}(P)
\end{array}
\right],
\end{equation}
where
\begin{eqnarray}
\label{piKEE}
{\cal K}_{EE}^\pi &= &\int_0^1d\alpha \left[ {\cal C}^{\rm iu}(\omega(M^2,\alpha,P^2))  - 2 \alpha(1-\alpha) \, P^2 \, \overline{\cal C}^{\rm iu}_1(\omega(M^2,\alpha,P^2))\right],\\
{\cal K}_{EF}^\pi &=& P^2 \int_0^1d\alpha\, \overline{\cal C}^{\rm iu}_1(\omega(M^2,\alpha,P^2)), \\
{\cal K}_{FE}^\pi &=& \frac{1}{2} M^2 \int_0^1d\alpha\, \overline{\cal C}^{\rm iu}_1(\omega(M^2,\alpha,P^2)),\\
{\cal K}_{FF}^\pi &=& - 2 {\cal K}_{FE}\,, \label{piKFF}
\end{eqnarray}
with $\overline{\cal C}_1(z) = {\cal C}_1(z)/z$.  We used Eq.\,(\ref{avwtiP}) to arrive at this form of ${\cal K}_{FF}$.  It follows immediately that the explicit form of Eq.\,(\ref{Gamma-qqscalar}) is:
\begin{equation}
\label{bsefinalqqscalar}
\left[
\begin{array}{c}
E_{qq_{0^+}}(P)\\
F_{qq_{0^+}}(P)
\end{array}
\right]
= \frac{1}{6\pi^2 m_G^2}
\left[
\begin{array}{cc}
{\cal K}_{EE}^\pi & {\cal K}_{EF}^\pi \\
{\cal K}_{FE}^\pi & {\cal K}_{FF}^\pi
\end{array}\right]
\left[\begin{array}{c}
E_{qq_{0^+}}(P)\\
F_{qq_{0^+}}(P)
\end{array}
\right].
\end{equation}
Equations~(\ref{bsefinal0}) and (\ref{bsefinalqqscalar}) are eigenvalue problems: they each have a solution at a single value of $P^2<0$, at which point the eigenvector describes the on-shell Bethe-Salpeter amplitude.

In the computation of observables, one must use the canonically-normalised Bethe-Salpeter amplitude; i.e., $\Gamma_\pi$ is rescaled so that
\begin{equation}
P_\mu = N_c\, {\rm tr} \int\! \frac{d^4q}{(2\pi)^4}\Gamma_\pi(-P)
 \frac{\partial}{\partial P_\mu} S(q+P) \, \Gamma_\pi(P)\, S(q)\,, \label{Ndefpion}
\end{equation}
where $N_c=3$.  For the pion in the chiral limit,  this expression assumes a particularly simple form; viz.,
\begin{equation}
1 = \frac{N_c}{4\pi^2} \frac{1}{M^2} \, {\cal C}_1(M^2;\tau_{\rm ir}^2,\tau_{\rm uv}^2)
E_\pi [ E_\pi - 2 F_\pi].
\label{Norm0}
\end{equation}
The canonical normalisation condition for the scalar diquark is almost identical to Eq.\,(\ref{Ndefpion}), with the only difference being the replacement $N_c = 3 \to 2$.

In order to estimate the mass of the first radial excitation of the pion, we use the following kernel
\begin{eqnarray}
\label{piKEEradial}
{\cal K}_{EE}^{\pi^\ast} &= &\int_0^1d\alpha \left[ {\cal F}^{\rm iu}(\omega(M^2,\alpha,P^2))  - 2 \alpha(1-\alpha) \, P^2 \, \overline{\cal F}^{\rm iu}_1(\omega(M^2,\alpha,P^2))\right],\\
{\cal K}_{EF}^{\pi^\ast} &=& P^2 \int_0^1d\alpha\, \overline{\cal F}^{\rm iu}_1(\omega(M^2,\alpha,P^2)), \\
{\cal K}_{FE}^{\pi^\ast} &=& \frac{1}{2} M^2 \int_0^1d\alpha\, \overline{\cal F}^{\rm iu}_1(\omega(M^2,\alpha,P^2)) -
\frac{1}{2} M_0^2 \int_0^1d\alpha\, \overline{\cal F}^{\rm iu}_1(\omega(M_0^2,\alpha,P^2)),\\
{\cal K}_{FF}^{\pi^\ast} &=& - 2 {\cal K}_{FE}\,. \label{piKFFradial}
\end{eqnarray}
It is conceived so that the leptonic decay constant of the radially-excited pseudoscalar meson vanishes in the chiral limit, which is a consequence of the axial-vector Ward-Takashi identity \cite{Holl:2004fr}.  The kernel for the radially-excited scalar diquark is obtained through obvious analogy with Eq.\,(\ref{bsefinalqqscalar}).

\subsection{Mesons and diquarks with \mbox{$J=1$}}
\label{app:vectors}
In the treatment of Eq.\,(\ref{njlgluon}) using the rainbow-ladder truncation the vector-meson Bethe-Salpeter amplitude has a particulary simple form; viz.,
\begin{equation}
\Gamma_\mu^{1^-}(P) =  \gamma_\mu^\perp E_{1^-}(P)\,,\; \gamma_\mu^\perp P_\mu=0.
\end{equation}
Hence the explicit form of Eq.\,(\ref{genbse}) for the ground-state vector-meson, whose solution yields its mass-squared, is
\begin{equation}
1+ K^\rho(-m_{1^-}^2) = 0\,,\;
K^\rho(P^2) = \frac{1}{3\pi^2m_G^2} \int_0^1d\alpha\, \alpha(1-\alpha) P^2\,  \overline{\cal C}^{iu}_1(\omega(M^2,\alpha,P^2))\,. \label{Kgamma}
\end{equation}
Equation~(\ref{avwtiP}) was used to express the BSE in this form.  The BSE for the axial-vector diquark again follows immediately; viz.,
\begin{equation}
1+ \frac{1}{2} K^\rho(-m_{qq_{1^+}}^2) = 0\,.
\end{equation}
The canonical normalisation conditions are readily expressed; viz.,
\begin{equation}
\label{CNavdiquark}
\frac{1}{E_{1^-}^2} = - \left.9 m_G^2 \frac{d}{dP^2} K^\rho(P^2)\right|_{P^2=-m_{1^-}^2},\;
\frac{1}{E_{qq_{1^+}}^2} = - \left.6 m_G^2 \frac{d}{dP^2} K^\rho(P^2)\right|_{P^2=-m_{qq_{1^+}}^2}\,.
\end{equation}

We emphasise that the vector-meson and axial-vector-diquark BSEs only assume such particularly simple forms in the rainbow-ladder truncation.  Even with a momentum-independent interaction, vector meson and axial-vector diquark Bethe-Salpeter amplitudes possess two Dirac covariants immediately upon inclusion of next-to-leading-order corrections to the quark-gluon vertex; viz.,
\begin{equation}
\Gamma_\mu^{1^-}(P) =  \gamma_\mu^\perp E_{1^-}(P) \;
\longrightarrow \;  \gamma_\mu^\perp E_{1^-}(P) + i\frac{1}{M} \sigma_{\mu\nu} P_\nu F_{1^-}(P)\,,\; \gamma^\perp_\mu P_\mu = 0.
\end{equation}
Similar observations hold for a $g^2 D(p-q) \sim \delta^4(p-q)$ interaction \cite{Bender:1996bb,Bender:2002as,Bhagwat:2004hn}.

Again owing to the simplicity of the interaction, the Bethe-Salpeter amplitude for an axial-vector meson is
\begin{equation}
\label{avBSA}
\Gamma_\mu^{1^+}(P) =  \gamma_5\gamma_\mu^\perp E_{1^+}(P)\,.
\end{equation}
In this case dressing the vertex does not generate new covariants because a momentum-independent interaction cannot generate a Bethe-Salpeter amplitude that depends on the relative momentum.  Inserting Eq.\,(\ref{avBSA}) into Eq.\,(\ref{LBSE}) yields the following BSE:
\begin{equation}
\label{a1Kernel}
1 + K^{a_1}(-m_{1^+}^2) = 0\,,\;
K^{a_1}(P^2) = - \frac{1}{3\pi^2m_G^2} \int_0^1d\alpha\, {\cal C}_1^{\rm iu}(\omega(M^2,\alpha,P^2))\,.
\end{equation}
It follows that the vector-diquark mass is determined by
\begin{equation}
1 + \frac{1}{2} K^{a_1}(-m_{qq_{1^-}}^2) = 0\,.
\end{equation}

The canonical normalisation conditions are
\begin{equation}
\frac{1}{E_{1^+}^2} = - \left.9 m_G^2 \frac{d}{dP^2} K^{a_1}(P^2)\right|_{P^2=-m_{1^+}^2},\;
\frac{1}{E_{qq_{1^-}}^2} = - \left.6 m_G^2 \frac{d}{dP^2} K^{a_1}(P^2)\right|_{P^2=-m_{qq_{1^-}}^2}\,.
\end{equation}

\subsection{Scalar mesons and pseudoscalar diquarks}
\label{app:scalar}
The Bethe-Salpeter amplitude for a scalar meson is
\begin{equation}
\label{sBSA}
\Gamma_{0^+}(P) = \mbox{\boldmath $I$}_{\rm D} \, E_{0^+}(P)\,.
\end{equation}
As with axial-vector mesons, dressing the vertex does not generate new covariants.  Inserting Eq.\,(\ref{sBSA}) into Eq.\,(\ref{LBSE}) yields the following BSE:
\begin{eqnarray}
\label{scalarBSEg}
1 &=&  - \frac{4}{3}\frac{1}{m_G^2} \int \! \frac{d^4q}{(2\pi)^4} \gamma_\mu S(q+P) S(q) \gamma_\mu \\
& = & \frac{16}{3}\frac{1}{m_G^2} \int \! \frac{d^4q}{(2\pi)^4} \frac{q^2+ q\cdot P - M^2}{[(q+P)^2+M^2][q^2+M^2]}\,. \label{scalarBSE}
\end{eqnarray}
Now consider Eq.\,(\ref{Mavwti}): if one sets $P^2=-4M^2$ in that chiral limit identity, then one finds after just two lines of algebra that it is equivalent to Eq.\,(\ref{scalarBSE}).  Hence, for $m=0$ the treatment of Eq.\,(\ref{njlgluon}) using the rainbow-ladder truncation yields \cite{Roberts:2010gh}
\begin{equation}
m_{0^+} = 2 \,M\,.
\end{equation}

For general values of the current-quark mass, using our symmetry-preserving regularisation prescription, Eq.\,(\ref{scalarBSEg}) can be written
\begin{equation}
\label{sigmaKernel}
1 + K^{\sigma}(-m_{0^+}^2) = 0\,,\;
K^{\sigma}(P^2) = \frac{1}{3\pi^2m_G^2} \int_0^1d\alpha\,
\left[{\cal C}^{\rm iu}(\omega(M^2,\alpha,P^2))-2 \,{\cal C}_1^{\rm iu}(\omega(M^2,\alpha,P^2))\right]\,.
\end{equation}
It follows that in the rainbow-ladder truncation the mass of a pseudoscalar diquark is determined by
\begin{equation}
1 + \frac{1}{2}K^{\sigma}(-m_{qq_{0^-}}^2) = 0\,.
\end{equation}

The canonical normalisation conditions are
\begin{equation}
\frac{1}{E_{0^+}^2} = - \left. \frac{9}{2} m_G^2 \frac{d}{dP^2} K^{\sigma}(P^2)\right|_{P^2=-m_{0^+}^2},\;
\frac{1}{E_{qq_{0^-}}^2} = - \left. 3 m_G^2 \frac{d}{dP^2} K^{\sigma}(P^2)\right|_{P^2=-m_{qq_{0^-}}^2}\,.
\end{equation}

\setcounter{equation}{0}
\section{Faddeev Equation}
\setcounter{dumbone}{\arabic{section}}
\label{app:FE}
\subsection{General structure}
\label{app:FEgeneral}
The nucleon is represented by a Faddeev amplitude
\begin{equation}
\label{PsiNucleon}
\Psi = \Psi_1 + \Psi_2 + \Psi_3  \,,
\end{equation}
where the subscript identifies the bystander quark and, e.g., $\Psi_{1,2}$ are obtained from $\Psi_3$ by a cyclic permutation of all the quark labels.  We employ the simplest realistic representation of $\Psi$.  The spin- and isospin-$1/2$ nucleon is a sum of scalar and axial-vector diquark correlations:
\begin{equation}
\label{Psi} \Psi_3(p_i,\alpha_i,\tau_i) = {\cal N}_3^{0^+} + {\cal N}_3^{1^+},
\end{equation}
with $(p_i,\alpha_i,\tau_i)$ the momentum, spin and isospin labels of the
quarks constituting the bound state, and $P=p_1+p_2+p_3$ the system's total momentum.

It is conceivable that pseudoscalar and vector diquarks could play a role in the ground-state nucleon's Faddeev amplitude.  However, they have parity opposite to that of the nucleon and hence can only appear in concert with nonzero quark angular momentum.  Since one expects the ground-state nucleon to possess the minimum possible amount of quark orbital angular momentum and these diquark correlations are significantly more massive than the scalar and axial-vector (Table~\ref{Diquarkmasses}), they can safely be ignored in computing properties of the ground state.

The scalar diquark piece in Eq.\,(\ref{Psi}) is
\begin{eqnarray}
{\cal N}_3^{0^+}(p_i,\alpha_i,\tau_i)&=& [\Gamma^{0^+}(\frac{1}{2}p_{[12]};K)]_{\alpha_1
\alpha_2}^{\tau_1 \tau_2}\, \Delta^{0^+}(K) \,[{\cal S}(\ell;P) u(P)]_{\alpha_3}^{\tau_3}\,,%
\label{calS}
\end{eqnarray}
where: the spinor satisfies (App.\,\protect\ref{App:EM})
\begin{equation}
(i\gamma\cdot P + M)\, u(P) =0= \bar u(P)\, (i\gamma\cdot P + M)\,,
\end{equation}
with $M$ the mass obtained by solving the Faddeev equation, and it is also a
spinor in isospin space with $\varphi_+= {\rm col}(1,0)$ for the proton and
$\varphi_-= {\rm col}(0,1)$ for the neutron; $K= p_1+p_2=: p_{\{12\}}$,
$p_{[12]}= p_1 - p_2$, $\ell := (-p_{\{12\}} + 2 p_3)/3$;
\begin{equation}
\Delta^{0^+}(K) = \frac{1}{K^2+m_{qq_{0^+}}^2}
\end{equation}
is a propagator for the scalar diquark formed from quarks $1$ and $2$, with $m_{0^+}$ the mass-scale associated with this correlation, and $\Gamma^{0^+}\!$ is the canonically-normalised Bethe-Salpeter amplitude describing their relative momentum correlation, Sect.\,\ref{app:groundBSEspion}; and ${\cal S}$, a $4\times 4$ Dirac matrix, describes the relative quark-diquark momentum correlation.  The colour antisymmetry of $\Psi_3$ is implicit in $\Gamma^{J^P}\!\!$, with the
Levi-Civita tensor, $\epsilon_{c_1 c_2 c_3}$, expressed via the antisymmetric
Gell-Mann matrices; viz., defining
\begin{equation}
\label{Hmatrices}
\{H^1=i\lambda^7,H^2=-i\lambda^5,H^3=i\lambda^2\}\,,\;
\mbox{then}\; \epsilon_{c_1 c_2 c_3}= (H^{c_3})_{c_1 c_2}.
\end{equation}

The axial-vector component in Eq.\,(\ref{Psi}) is
\begin{eqnarray}
{\cal N}^{1^+}(p_i,\alpha_i,\tau_i) & =&  [{\tt t}^i\,\Gamma_\mu^{1^+}(\frac{1}{2}p_{[12]};K)]_{\alpha_1
\alpha_2}^{\tau_1 \tau_2}\,\Delta_{\mu\nu}^{1^+}(K)\,
[{\cal A}^{i}_\nu(\ell;P) u(P)]_{\alpha_3}^{\tau_3}\,,
\label{calA}
\end{eqnarray}
where the symmetric isospin-triplet matrices are
\begin{equation}
{\tt t}^+ = \frac{1}{\surd 2}(\tau^0+\tau^3) \,,\;
{\tt t}^0 = \tau^1\,,\;
{\tt t}^- = \frac{1}{\surd 2}(\tau^0-\tau^3)\,,
\end{equation}
and the other elements in Eq.\,(\ref{calA}) are straightforward generalisations of those in Eq.\,(\ref{calS}) with, e.g.,
\begin{equation}
\Delta_{\mu\nu}^{1^+}(K) = \frac{1}{K^2+m_{qq_{1^+}}^2} \, \left(\delta_{\mu\nu} + \frac{K_\mu K_\nu}{m_{qq_{1^+}}^2}\right) \,.
\end{equation}

Since it is not possible to combine an isospin-0 diquark with an isospin-1/2 quark to obtain isospin-3/2, the spin- and isospin-3/2 $\Delta$ contains only an axial-vector diquark component
\begin{equation}
\Psi_3^\Delta(p_i,\alpha_i,\tau_i) = {\cal D}_3^{1+}.
\end{equation}
Understanding the structure of the $\Delta$ is plainly far simpler than in the case of the nucleon since, whilst the general form of the Faddeev amplitude for a spin- and isospin-3/2 can be complicated, isospin symmetry means that one can focus on the $\Delta^{++}$, with its simple flavour structure, because all the charge states are degenerate:
\begin{equation}
{\cal D}_3^{1^+}= [{\tt t}^+ \Gamma^{1^+}_\mu(\frac{1}{2}p_{[12]};K)]_{\alpha_1 \alpha_2}^{\tau_1 \tau_2}
\, \Delta_{\mu\nu}^{1^+}(K) \, [{\cal D}_{\nu\rho}(\ell;P)u_\rho(P)\, \varphi_+]_{\alpha_3}^{\tau_3}\,, \label{DeltaAmpA}
\end{equation}
where $u_\rho(P)$ is a Rarita-Schwinger spinor, Eq.\,(\ref{rarita}).

The general forms of the matrices ${\cal S}(\ell;P)$, ${\cal A}^i_\nu(\ell;P)$ and ${\cal D}_{\nu\rho}(\ell;P)$, which describe the momentum-space correlation between the quark and diquark in the nucleon and $\Delta$, respectively, are described in Refs.\,\cite{Cloet:2007pi,Oettel:1998bk}.  The requirement that ${\cal S}(\ell;P)$ represent a positive energy nucleon entails
\begin{equation}
\label{Sexp}
{\cal S}(\ell;P) = s_1(\ell;P)\,\mbox{\boldmath $I$}_{\rm D} + \left(i\gamma\cdot \hat\ell - \hat\ell \cdot \hat P\, \mbox{\boldmath $I$}_{\rm D}\right)\,s_2(\ell;P)\,,
\end{equation}
where $(\mbox{\boldmath $I$}_{\rm D})_{rs}= \delta_{rs}$, $\hat \ell^2=1$, $\hat P^2= - 1$.  In the nucleon rest frame, $s_{1,2}$ describe, respectively, the upper, lower component of the bound-state nucleon's spinor.  Placing the same constraint on the axial-vector component, one has
\begin{equation}
\label{Aexp}
 {\cal A}^i_\nu(\ell;P) = \sum_{n=1}^6 \, p_n^i(\ell;P)\,\gamma_5\,A^n_{\nu}(\ell;P)\,,\; i=+,0,-\,,
\end{equation}
where ($ \hat \ell^\perp_\nu = \hat \ell_\nu + \hat \ell\cdot\hat P\, \hat P_\nu$, $ \gamma^\perp_\nu = \gamma_\nu + \gamma\cdot\hat P\, \hat P_\nu$)
\begin{equation}
\label{Afunctions}
\begin{array}{lll}
A^1_\nu= \gamma\cdot \hat \ell^\perp\, \hat P_\nu \,,\; &
A^2_\nu= -i \hat P_\nu \,,\; &
A^3_\nu= \gamma\cdot\hat \ell^\perp\,\hat \ell^\perp\,,\\
A^4_\nu= i \,\hat \ell_\mu^\perp\,,\; &
A^5_\nu= \gamma^\perp_\nu - A^3_\nu \,,\; &
A^6_\nu= i \gamma^\perp_\nu \gamma\cdot\hat \ell^\perp - A^4_\nu\,.
\end{array}
\end{equation}
Finally, requiring also that ${\cal D}_{\nu\rho}(\ell;P)$ be an eigenfunction of $\Lambda_+(P)$, one obtains
\begin{equation}
\label{DeltaFA}
{\cal D}_{\nu\rho}(\ell;P) = {\cal S}^\Delta(\ell;P) \, \delta_{\nu\rho} + \gamma_5{\cal A}_\nu^\Delta(\ell;P) \,\ell^\perp_\rho \,,
\end{equation}
with ${\cal S}^\Delta$ and ${\cal A}^\Delta_\nu$ given by obvious analogues of Eqs.\,(\ref{Sexp}) and (\ref{Aexp}), respectively.

One can now write the Faddeev equation satisfied by $\Psi_3$ as
\begin{equation}
 \left[ \begin{array}{r}
{\cal S}(k;P)\, u(P)\\
{\cal A}^i_\mu(k;P)\, u(P)
\end{array}\right]\\
 = -\,4\,\int\frac{d^4\ell}{(2\pi)^4}\,{\cal M}(k,\ell;P)
\left[
\begin{array}{r}
{\cal S}(\ell;P)\, u(P)\\
{\cal A}^j_\nu(\ell;P)\, u(P)
\end{array}\right] .
\label{FEone}
\end{equation}
The kernel in Eq.~(\ref{FEone}) is
\begin{equation}
\label{calM} {\cal M}(k,\ell;P) = \left[\begin{array}{cc}
{\cal M}_{00} & ({\cal M}_{01})^j_\nu \\
({\cal M}_{10})^i_\mu & ({\cal M}_{11})^{ij}_{\mu\nu}\rule{0mm}{3ex}
\end{array}
\right] ,
\end{equation}
with
\begin{equation}
 {\cal M}_{00} = \Gamma^{0^+}\!(k_q-\ell_{qq}/2;\ell_{qq})\,
S^{\rm T}(\ell_{qq}-k_q) \,\bar\Gamma^{0^+}\!(\ell_q-k_{qq}/2;-k_{qq})\,
S(\ell_q)\,\Delta^{0^+}(\ell_{qq}) \,,
\end{equation}
where: $\ell_q=\ell+P/3$, $k_q=k+P/3$, $\ell_{qq}=-\ell+ 2P/3$,
$k_{qq}=-k+2P/3$ and the superscript ``T'' denotes matrix transpose; and
\begin{eqnarray}
({\cal M}_{01})^j_\nu &=& {\tt t}^j \,
\Gamma_\mu^{1^+}\!(k_q-\ell_{qq}/2;\ell_{qq}) S^{\rm T}(\ell_{qq}-k_q)\,\bar\Gamma^{0^+}\!(\ell_q-k_{qq}/2;-k_{qq})\,
S(\ell_q)\,\Delta^{1^+}_{\mu\nu}(\ell_{qq}) \,, \label{calM01} \\
({\cal M}_{10})^i_\mu &=&
\Gamma^{0^+}\!(k_q-\ell_{qq}/2;\ell_{qq})\,
S^{\rm T}(\ell_{qq}-k_q)\,{\tt t}^i\, \bar\Gamma_\mu^{1^+}\!(\ell_q-k_{qq}/2;-k_{qq})\,
S(\ell_q)\,\Delta^{0^+}(\ell_{qq}) \,,\\
({\cal M}_{11})^{ij}_{\mu\nu} &=& {\tt t}^j\,
\Gamma_\rho^{1^+}\!(k_q-\ell_{qq}/2;\ell_{qq})\, S^{\rm T}(\ell_{qq}-k_q)\,{\tt t}^i\, \bar\Gamma^{1^+}_\mu\!(\ell_q-k_{qq}/2;-k_{qq})\,
S(\ell_q)\,\Delta^{1^+}_{\rho\nu}(\ell_{qq}) \,. \label{calM11}
\end{eqnarray}

The $\Delta$'s Faddeev equation is
\begin{eqnarray}
{\cal D}_{\lambda\rho}(k;P)\,u_\rho(P) & = & 4\int\frac{d^4\ell}{(2\pi)^4}\,{\cal
M}^\Delta_{\lambda\mu}(k,\ell;P) \,{\cal D}_{\mu\sigma}(\ell;P)\,u_\sigma(P)\,, \label{FEDelta}
\end{eqnarray}
with
\begin{equation}
{\cal M}^\Delta_{\lambda\mu} = {\tt t}^+
\Gamma_\sigma^{1^+}\!(k_q-\ell_{qq}/2;\ell_{qq})\,
 S^{\rm T}\!(\ell_{qq}-k_q)\, {\tt t}^+\bar\Gamma^{1^+}_\lambda\!(\ell_q-k_{qq}/2;-k_{qq})\,
S(\ell_q)\,\Delta^{1^+}_{\sigma\mu}\!(\ell_{qq}).
\end{equation}

\subsection{Explicit form of the nucleon's Faddeev equation}
\label{app:FEnucleon}
Using Eq.\,(\ref{staticexchange}), the nucleon's Faddeev amplitude simplifies and can be written in terms of, Eqs.\,(\ref{Sexp}), (\ref{Aexp}),
\begin{equation}
{\cal S}(P) = s(P) \,\mbox{\boldmath $I$}_{\rm D}\,,\;
{\cal A}^i_\mu(P) = a_1^i(P) \gamma_5\gamma_\mu + a_2^i(P) \gamma_5 \hat P_\mu \,,i=+,0\,.
\end{equation}
The mass of the ground-state nucleon is then determined by a $5\times 5$ matrix Faddeev equation; viz.,
\begin{equation}
\left(\begin{array}{c}
s(P)\\[0.7ex]
a_1^+(P)\\[0.7ex]
a_1^0(P)\\[0.7ex]
a_2^+(P)\\[0.7ex]
a_2^0(P)\end{array}\right)
=
\left( \begin{array}{ccccc}
K^{00}_{ss} & -\surd 2 \, K^{01}_{sa_1} & K^{01}_{sa_1} & -\surd 2\, K^{01}_{sa_2} & K^{01}_{sa_2}\\[0.7ex]
-\surd 2\, K^{10}_{a_1 s} & 0 & \surd 2\, K^{11}_{a_1 a_1} & 0 & \surd 2\, K^{11}_{a_1 a_2}\\[0.7ex]
K^{10}_{a_1 s} & \surd 2 \, K^{11}_{a_1 a_1} & K^{11}_{a_1 a_1} & \surd 2\,K^{11}_{a_1 a_2} & K^{11}_{a_1 a_2} \\[0.7ex]
-\surd 2\, K^{10}_{a_2 s} & 0 & \surd 2\, K^{11}_{a_2 a_1} & 0 & \surd 2\, K^{11}_{a_2 a_2} \\[0.7ex]
K^{10}_{a_2 s} & \surd 2\, K^{11}_{a_2 a_1} & K^{11}_{a_2 a_1} & \surd 2\, K^{11}_{a_2 a_2} & K^{11}_{a_2 a_2}
\end{array}
\right)
\left(\begin{array}{c}
s(P)\\[0.7ex]
a_1^+(P)\\[0.7ex]
a_1^0(P)\\[0.7ex]
a_2^+(P)\\[0.7ex]
a_2^0(P)\end{array}\right)
\end{equation}
where: $c_N = g_N^2/(4 \pi^2 M)$,
\begin{equation}
\sigma_N^0 = \sigma_N(\alpha,M,m_{qq_{0^+}},m_N) := (1-\alpha)\,M^2 + \alpha\,m_{qq_{0^+}}^2 - \alpha (1-\alpha) m_N^2\,,\;
\sigma_N^1 = \sigma_N(\alpha,M,m_{qq_{1^+}},m_N) \,;
\end{equation}
and
\begin{eqnarray}
K^{00}_{ss} & = & K^{00}_{EE}+K^{00}_{EF}+K^{00}_{FF}\,,\\
K^{00}_{EE} & = & c_N E_{qq_{0^+}}^2 \!
\int_0^1 d\alpha \,\overline{\cal C}_1(\sigma_N^0)
(\alpha m_N + M)\,,\\
K^{00}_{EF} & = & - 2 c_N E_{qq_{0^+}} F_{qq_{0^+}}\frac{m_N}{M} \!
\int_0^1 d\alpha \,\overline{\cal C}_1(\sigma_N^0)
(1-\alpha) (\alpha m_N + M)\,,\\
K^{00}_{FF} & = & c_N F_{qq_{0^+}}^2\frac{m_{qq_{0^+}}^2}{M^2} \!
\int_0^1 d\alpha \,\overline{\cal C}_1(\sigma_N^0)(\alpha m_N + M)\,;\\
K^{01}_{s a_1} & = & K^{01}_{s_E a_1} + K^{10}_{s_F a_1}\,,\\
K^{01}_{s_E a_1} &=& c_N \frac{E_{qq_{0^+}}E_{qq_{1^+}}}{m_{qq_{1^+}}^2}\!
\int_0^1 d\alpha \,\overline{\cal C}_1(\sigma_N^1)
( m_{qq_{1^+}}^2 (3 M + \alpha m_N) + 2 \alpha (1-\alpha)^2 m_N^3 )\,,\\
K^{01}_{s_F a_1} &=& -c_N \frac{F_{qq_{0^+}}E_{qq_{1^+}}}{m_{qq_{1^+}}^2} \frac{m_N}{M}\!
\int_0^1 d\alpha \,\overline{\cal C}_1(\sigma_N^1)
(1-\alpha)
(m_{qq_{1^+}}^2 (M + 3 \alpha m_N) + 2 (1-\alpha)^2 M m_N^2) \,;\\
K^{01}_{s a_2} & = & K^{01}_{s_E a_2} + K^{01}_{s_F a_2}\,,\\
K^{01}_{s_E a_2} & = & c_N \frac{E_{qq_{0^+}}E_{qq_{1^+}}}{m_{qq_{1^+}}^2}\!
\int_0^1 d\alpha \,\overline{\cal C}_1(\sigma_N^1)
(\alpha m_N - M) ((1-\alpha)^2 m_N^2-m_{qq_{1^+}}^2)\,,\\
K^{01}_{s_F a_2} & = & c_N \frac{F_{qq_{0^+}}E_{qq_{1^+}}}{m_{qq_{1^+}}^2} \frac{m_N}{M} \!
\int_0^1 d\alpha \,\overline{\cal C}_1(\sigma_N^1)
(1-\alpha)(\alpha m_N - M) ((1-\alpha)^2 m_N^2-m_{qq_{1^+}}^2)\,;\\
K^{10}_{a_1 s} & = & K^{10}_{a_1 s_E} + K^{10}_{a_1 s_F}\,,\\
K^{10}_{a_1 s_E} & = & \frac{c_N}{3}\frac{E_{qq_{0^+}}E_{qq_{1^+}}}{m_{qq_{1^+}}^2} \!
\int_0^1 d\alpha \,\overline{\cal C}_1(\sigma_N^0)
(\alpha m_N + M) (2 m_{qq_{1^+}}^2 + (1-\alpha)^2 m_N^2)\,,\\
K^{10}_{a_1 s_F} & = & -\frac{c_N}{3}\frac{F_{qq_{0^+}}E_{qq_{1^+}}}{m_{qq_{1^+}}^2} \frac{m_N}{M} \!
\int_0^1 d\alpha \,\overline{\cal C}_1(\sigma_N^0)
(1-\alpha)(2 m_{qq_{1^+}}^2 + (1-\alpha)^2 m_N^2) (\alpha m_N + M)\,;\\
K^{10}_{a_2 s} & = & K^{10}_{a_2 s_E} + K^{10}_{a_2 s_F}\,,\\
K^{10}_{a_2 s_E} & = & \frac{c_N}{3} \frac{E_{qq_{0^+}}E_{qq_{1^+}}}{m_{qq_{1^+}}^2} \!
\int_0^1 d\alpha \,\overline{\cal C}_1(\sigma_N^0)
(\alpha m_N + M) (m_{qq_{1^+}}^2 - 4 (1-\alpha)^2 m_N^2),\\
K^{10}_{a_2 s_F} & = & \frac{c_N}{3} \frac{F_{qq_{0^+}}E_{qq_{1^+}}}{m_{qq_{1^+}}^2}\frac{m_N}{M} \!
\int_0^1 d\alpha \,\overline{\cal C}_1(\sigma_N^0)
(1-\alpha) (5 m_{qq_{1^+}}^2-2(1-\alpha)^2 m_N^2)(\alpha m_N + M)\,;\\
K^{11}_{a_1 a_1} & = & -\frac{c_N}{3}\frac{E_{qq_{1^+}}^2}{m_{qq_{1^+}}^2} \!
\int_0^1 d\alpha \,\overline{\cal C}_1(\sigma_N^1)
[ 2 m_{qq_{1^+}}^2 (M-\alpha m_N) + (1-\alpha)^2 m_N^2 (M+5 \alpha m_N)]\,;\\
K^{11}_{a_1 a_2} & = & -\frac{2 c_N}{3}\frac{E_{qq_{1^+}}^2}{m_{qq_{1^+}}^2} \!
\int_0^1 d\alpha \,\overline{\cal C}_1(\sigma_N^1)
(-m_{qq_{1^+}}^2+(1-\alpha^2) m_N^2) (\alpha m_N - M)\,;\\
K^{11}_{a_2 a_1} & = & -\frac{c_N}{3}\frac{E_{qq_{1^+}}^2}{m_{qq_{1^+}}^2} \!
\int_0^1 d\alpha \,\overline{\cal C}_1(\sigma_N^1)
[m_{qq_{1^+}}^2(11 \alpha  m_N + M) - 2(1-\alpha)^2 m_N^2 (7\alpha m_N + 2 M)]\,;\\
K^{11}_{a_2 a_2}  & = & - \frac{5 c_N}{3} \frac{E_{qq_{1^+}}^2}{m_{qq_{1^+}^2}} \!
\int_0^1 d\alpha \,\overline{\cal C}_1(\sigma_N^1)
(m_{qq_{1^+}}^2 - (1-\alpha)^2 m_N^2) (\alpha m_N - M)\,.
\end{eqnarray}
This kernel was computed following the procedure detailed for the $\Delta$-resonance in Sect.\,\ref{DeltaFEexplicit}.  During this process we employed the replacements in Eq.\,(\ref{replacementsDelta}), their analogues involving the scalar-diquark's momentum, $K_{0^+}$, and $K_{0^+}\cdot K_{1^+}\to (1-\alpha)^2 P^2$.  In the present context, of course, $P^2=-m_N^2$.

Given the structure of the kernel, it is not surprising that the eigenvectors exhibit the pattern
\begin{equation}
a_i^+ = -\surd 2 a_i^0,\, i=1,2\,.
\end{equation}
For example, at the mass presented in Table~\ref{Table:baryon}, the nucleon's unit-normalised Faddeev amplitude is
\begin{equation}
\begin{array}{ccccc}
s & a_1^+ & a_1^0 & a_2^+ & a_2^0 \\
0.88 & 0.38 & -0.27 & -0.065 & 0.046
\end{array}\,.
\end{equation}
The axial-vector-diquark correlation provides 22\% of the unit normalisation.  This is discussed further in connection with Fig.\,\ref{Fig6}.




\end{document}